\documentclass[journal=jacsat,manuscript=article]{achemso}

\usepackage[version=3]{mhchem} % Formula subscripts using \ce{}
\usepackage{graphicx}% Include figure files
\usepackage{dcolumn}% Align table columns on decimal point
\usepackage{bm}% bold math
\usepackage{amssymb,amsfonts}
\usepackage{amsmath}
\usepackage{subcaption}
\usepackage{longtable}
\usepackage{units}
\usepackage{xcolor}
\usepackage{braket}
\newcommand{\highlight}[1]{{\color{red} #1}}

\newcommand{\ReSpect}{\textsc{ReSpect}}

\DeclareMathOperator{\Tr}{Tr}

\renewcommand{\vec}[1]{\boldsymbol{#1}}
\newcommand{\mat}[1]{\mathbf{#1}}

\author{Michal Repisky}
\email{michal.repisky@uit.no}
\affiliation{%
 Hylleraas Centre for Quantum Molecular Sciences, Department of Chemistry, UiT The Arctic University of Norway, N-9037 Tromsø, Norway
}%
\alsoaffiliation{%
 Department of Physical and Theoretical Chemistry, Faculty of Natural Sciences, Comenius University, SK-84215 Bratislava, Slovakia 
}%

\author{Stanislav Komorovsky}%
\email{stanislav.komorovsky@savba.sk}
\affiliation{%
 Institute of Inorganic Chemistry, Slovak Academy of Sciences, Dubravska cesta 9, SK-84536 Bratislava, Slovakia
}%

\author{Lukas Konecny}
\affiliation{%
 Hylleraas Centre for Quantum Molecular Sciences, Department of Chemistry, UiT The Arctic University of Norway, N-9037 Tromsø, Norway
}%
\alsoaffiliation{%
Max Planck Institute for the Structure and Dynamics of Matter, Center for Free Electron Laser Science, Luruper Chaussee 149, 22761 Hamburg, Germany
}%
\alsoaffiliation{%
 Department of Inorganic Chemistry, Faculty of Natural Sciences, Comenius University, SK-84215 Bratislava, Slovakia 
}%

\author{Marius Kadek}
\affiliation{%
 Hylleraas Centre for Quantum Molecular Sciences, Department of Chemistry, UiT The Arctic University of Norway, N-9037 Tromsø, Norway
}%

\author{Torsha Moitra}
\affiliation{%
 Department of Physical and Theoretical Chemistry, Faculty of Natural Sciences, Comenius University, SK-84215 Bratislava, Slovakia
}%
\alsoaffiliation{%
 Hylleraas Centre for Quantum Molecular Sciences, Department of Chemistry, UiT The Arctic University of Norway, N-9037 Tromsø, Norway
}%

\author{Marc Joosten}
\affiliation{%
 Hylleraas Centre for Quantum Molecular Sciences, Department of Chemistry, UiT The Arctic University of Norway, N-9037 Tromsø, Norway
}%

\author{Debora Misenkova}
\affiliation{%
 Institute of Inorganic Chemistry, Slovak Academy of Sciences, Dubravska cesta 9, SK-84536 Bratislava, Slovakia
}%

\author{Rasmus Vikhamar-Sandberg}
\affiliation{%
 Hylleraas Centre for Quantum Molecular Sciences, Department of Chemistry, UiT The Arctic University of Norway, N-9037 Tromsø, Norway
}%

\author{Martin Kaupp}
\affiliation{%
Technische Universit\"at Berlin, Institute of Chemistry, Strasse des 17 Juni 135, D-10623 Berlin, Germany
}%

\author{Kenneth Ruud}
\affiliation{%
 Norwegian Defence Research Establishment, P.O. Box 25, 2027 Kjeller, Norway
}%
\alsoaffiliation{%
 Hylleraas Centre for Quantum Molecular Sciences, Department of Chemistry, UiT The Arctic University of Norway, N-9037 Tromsø, Norway
}%

\author{Olga L. Malkina}
\affiliation{%
 Institute of Inorganic Chemistry, Slovak Academy of Sciences, Dubravska cesta 9, SK-84536 Bratislava, Slovakia
}%

\author{Vladimir G. Malkin}
\affiliation{%
 Institute of Inorganic Chemistry, Slovak Academy of Sciences, Dubravska cesta 9, SK-84536 Bratislava, Slovakia
}%

\title{X2C Hamiltonian Models in \ReSpect{}: Bridging Accuracy and Efficiency}

\begin{document}

%%%%%%%%%%%%%%%%%%%%%%%%%%%%%%%%%%%%%%%%%%%%%%%%%%%%%%%%%%%%%%%%%%%%%
%% The "tocentry" environment can be used to create an entry for the
%% graphical table of contents. It is given here as some journals
%% require that it is printed as part of the abstract page. It will
%% be automatically moved as appropriate.
%%%%%%%%%%%%%%%%%%%%%%%%%%%%%%%%%%%%%%%%%%%%%%%%%%%%%%%%%%%%%%%%%%%%%
%\begin{tocentry}
%
%Some journals require a graphical entry for the Table of Contents.
%This should be laid out ``print ready'' so that the sizing of the
%text is correct.

%\end{tocentry}

%%%%%%%%%%%%%%%%%%%%%%%%%%%%%%%%%%%%%%%%%%%%%%%%%%%%%%%%%%%%%%%%%%%%%
%% The abstract environment will automatically gobble the contents
%% if an abstract is not used by the target journal.
%%%%%%%%%%%%%%%%%%%%%%%%%%%%%%%%%%%%%%%%%%%%%%%%%%%%%%%%%%%%%%%%%%%%%
\begin{abstract}
Since its inception, the \ReSpect{} program has been evolving to provide powerful tools for simulating spectroscopic processes and exploring emerging research areas, all while incorporating relativistic effects, particularly spin–orbit interactions, in a fully variational manner. Recent developments have focused on exact two-component (X2C) Hamiltonian models that go beyond the standard one-electron X2C approach by incorporating two-electron picture-change corrections. This paper presents the theoretical foundations of two distinct atomic mean-field X2C models, amfX2C and extended eamfX2C, which offer computationally efficient and accurate alternatives to fully relativistic four-component methods. These models enable simulations of complex phenomena, such as time-resolved pump–probe spectroscopies and cavity-modified molecular properties, which would otherwise be computationally prohibitive. \ReSpect{} continues to evolve, providing state-of-the-art quantum chemical methods and post-processing tools, all available free of charge through our website, www.respectprogram.org, to support researchers exploring relativistic effects across various scientific disciplines.
\end{abstract}

%%%%%%%%%%%%%%%%%%%%%%%%%%%%%%%%%%%%%%%%%%%%%%%%%%%%%%%%%%%%%%%%%%%%%
%% Start the main part of the manuscript here.
%%%%%%%%%%%%%%%%%%%%%%%%%%%%%%%%%%%%%%%%%%%%%%%%%%%%%%%%%%%%%%%%%%%%%
\section{\label{sec:introduction}Introduction}

The behavior of electrons in atoms, molecules, and solids is fundamentally governed by the Dirac equation~\cite{Dirac1928,Dirac1928a}. Nevertheless, most quantum chemical calculations rely instead on the Schrödinger equation~\cite{Schrodinger1926}, which effectively assumes an infinite speed of light. The discrepancies between the solutions of the Dirac and Schrödinger equations are commonly referred to as relativistic effects. 

While often treated as minor corrections to the Schrödinger framework, relativistic effects can be crucial--even for systems with lighter elements~\cite{Pyykko1988}. For instance, in X-ray spectroscopy, spin-orbit (SO) interactions lift the degeneracy of atomic $p$- and $d$-orbitals, leading to the characteristic $L_{2,3}$, $M_{2,3}$, and $M_{4,5}$ edges~\cite{deGroot2005}. SO effects are also essential for interpreting the absorption spectra of lanthanide-based compounds used in medical imaging and photosensitization~\cite{Ning2019}. In Nuclear Magnetic Resonance (NMR) spectroscopy, relativistic corrections, such as those affecting proton shielding constants, can dominate over nonrelativistic contributions~\cite{Vicha2018}. In the solid state, SO interactions are central to spintronics~\cite{Wolf2001,Moore2010} and topological insulators~\cite{Hasan2010,Kou2017}.

The ubiquity of relativistic effects, coupled with the growing interest in heavy-element compounds, demands the development of efficient electronic structure methods that offer a variational treatment of relativistic phenomena—particularly spin–orbit interactions. The \ReSpect{} program was developed to address this need, enabling density functional theory (DFT) simulations of spectroscopic properties at the relativistic two- and four-component levels, including spin polarization in open-shell systems through a Kramers-unrestricted formalism. The spectroscopic properties currently available in \ReSpect{} are listed in Table~\ref{tab:overview}. Our previous publication detailed the core theoretical and technical foundations of the program based on the four-component formalism, with particular emphasis on the use of time-reversal symmetry and biquaternion algebra~\cite{Repisky2020}. These innovations substantially reduce computational cost and complexity, allowing four-component DFT calculations on systems with over 100 atoms using standard CPU-based clusters~\cite{Hrda2014,Gohr2015,Vicha2016-1,Repisky2020}, with computational overheads typically within an order of magnitude of non-relativistic methods.

Although the four-component (4c) formalism is regarded as the gold standard in relativistic quantum chemistry, performing 4c calculations on systems consisting of several hundred atoms, particularly those involving multiple heavy elements, is computationally demanding and challenging. As a result, researchers have sought approximate two-component (2c) Hamiltonians as a more efficient alternative. The primary advantage of 2c methods is that they simplify the problem by discarding negative-energy states (along with the two-electron integrals over the small-component basis associated with these states), effectively reducing the complexity of the original 4c approach by half. One 2c Hamiltonian that has gained significant popularity in recent years is the exact two-component (X2C) Hamiltonian~\cite{Heully1986,Jensen2005,Kutzelnigg2005,Liu2007,Ilias2007}. It reduces the 4c problem to a 2c one through straightforward algebraic manipulations, thus eliminating the need to explicitly calculate higher-order relativistic corrections and/or property operators.

There are several variants of the X2C Hamiltonian, each differing in the choice of the parent 4c Hamiltonian used to construct the 2c model~\cite{Jensen2005,Kutzelnigg2005,Liu2007,Ilias2007,Sikkema2009,Peng2007,Liu2009,Peng2013,Filatov2013,Konecny2016,Goings2016,Liu2018,Knecht2022,Zhang2022,Ehrman2023}. The one-electron X2C (1eX2C) model uses a pure one-electron Dirac Hamiltonian as the parent, where two-electron interactions are completely omitted from the X2C decoupling transformation step~\cite{Kutzelnigg2005,Liu2007,Ilias2007}. In contrast, the molecular mean-field X2C (mmfX2C) approach involves performing the X2C decoupling after completing a converged 4c molecular self-consistent field (SCF) calculation~\cite{Sikkema2009}. This method is typically applied in post-SCF electron correlation or property calculations. Between the 1eX2C and mmfX2C models, several parent Hamiltonian variants exist, which extend the 1eX2C approach by approximately including two-electron interactions~\cite{Ilias2007,Peng2007,Liu2009,Peng2013,Filatov2013,Konecny2016,Goings2016,Liu2018,Knecht2022,Zhang2022,Ehrman2023}. All of these models can be regarded as extensions or refinements of earlier conceptual frameworks: (i) element- and angular-momentum-specific screening factors in the evaluation of 1e spin-orbit (SO) integrals~\cite{Blume1962,Blume1963}, (ii) a mean-field SO approach~\cite{Hess1996} that forms the basis for the widely used AMFI module~\cite{AMFI}, and (iii) a method utilizing atomic model densities derived from Kohn-Sham density functional theory~\cite{Wullen2005}. The screening factors in (i) are sometimes referred to as “Boettger factors” or as the screened-nuclear-spin-orbit (SNSO) approach~\cite{Boettger2000}, which is originally derived from a second-order Douglas--Kroll--Hess DFT-based model. A later reparameterization of this method based on atomic four-component Dirac--Hartree--Fock results led to the modified SNSO (mSNSO) approach~\cite{Filatov2013}.

As part of ongoing efforts to advance the X2C framework, the \ReSpect{} team--together with Stefan Knecht, Hans Jørgen Aagaard Jensen, and Trond Saue from the DIRAC program~\cite{Saue2020}--recently developed and implemented two simple, computationally efficient, and numerically accurate X2C models: the atomic mean-field (amfX2C) and the extended atomic mean-field (eamfX2C)~\cite{Knecht2022}. These models expand on earlier work by Liu and Cheng~\cite{Liu2018}, incorporating full SO and scalar-relativistic corrections arising from two-electron interactions, whether they come from the Coulomb, Coulomb--Gaunt, or Coulomb--Breit Hamiltonian. Additionally, these approaches account for the characteristics of the underlying correlation framework (e.g., wave function theory or KS-DFT), allowing for the inclusion of tailor-made exchange-correlation corrections. Both X2C models have also been extended to property calculations, employing either response theory or a real-time approach~\cite{Konecny2023,Moitra2023}.

The present manuscript summarizes the theoretical foundations of two X2C Hamiltonian models as implemented in \ReSpect{}. The theoretical framework and representative applications of these models are presented in the context of self-consistent field (SCF) procedures, electron paramagnetic resonance (EPR), and time-resolved (TR) pump–probe spectroscopies, including electronic absorption (TR-EAS) and electronic circular dichroism (TR-ECD), using real-time TDDFT. Additionally, we discuss the extension of these models to the linear-response regime of TDDFT, incorporating explicit light–matter coupling via quantum electrodynamical density functional theory (QEDFT).

% \begin{table}[ht]
% \caption{
%    List of properties implemented in ReSpect, alongside
%    the specification of Hamiltonians, theoretical methods,
%    Kramers-restricted/unrestricted formalisms, and literature references.
% }
% \label{tab:Introduction}
% \begin{tabular}{lccccc}
%   \hline
%   \hline
%   Property & Hamiltonian & ~Method~ & ~KR~ & ~KU~  & Ref.\\
%   \hline
%   \hline
%   \\[-3pt]
%   \multicolumn{6}{c}{{\bf Self-Consistent Field (SCF)}} \\
%   \hline
%   Molecular          &  1c,2c,4c &     & \checkmark & \checkmark &    \\
%   Solid-state        &  1c,4c    &     & \checkmark &            &  \citenum{Kadek2019} \\[5pt]
% %
% % \hline
% \hline
% \end{tabular}
% \vspace*{1pt}

% Abbreviations in alphabetical order:
% %
% DR: damped response TDDFT,
% ZFS: zero-field splitting.
% %\end{flushleft}
% \end{table}

\renewcommand{\arraystretch}{0.6}
\begin{longtable}{lccccc}
\caption{
   List of properties implemented in \ReSpect{} as of 2025, along with the corresponding Hamiltonians, theoretical methods, Kramers-restricted (KR) and Kramers-unrestricted (KU) formalisms, and relevant literature references.
}
\label{tab:overview} \\
\hline\hline
Property & Hamiltonian & Method & KR & KU & Ref. \\
\hline\hline
\endfirsthead

\hline\hline
Property & Hamiltonian & Method & KR & KU & Ref. \\
\hline\hline
\endhead

\hline\hline
\multicolumn{6}{r}{\emph{Continued on next page}}\\
\endfoot

\hline\hline
\endlastfoot

\\[-3pt]
\multicolumn{6}{c}{\textbf{Self-Consistent Field (SCF)}}\\
\hline
Molecular          & 1c,2c,4c &     & \checkmark & \checkmark & \citenum{Repisky2020}   \\
Solid-state        & 1c,4c    &     & \checkmark &            & \citenum{Kadek2019,Kadek2023} \\[8pt]

\multicolumn{6}{c}{\textbf{Electron Paramagnetic Resonance (EPR)}}\\
\hline
ZFS                & 2c,4c    & EV  & - & \checkmark & \citenum{Repisky2020} \\
$g$-tensor         & 2c,4c    & PT1 & - & \checkmark & \citenum{Repisky2010,Gohr2015,Misenkova2022} \\
$A$-tensor         & 2c,4c    & PT1 & - & \checkmark & \citenum{Malkin2011,Gohr2015} \\[8pt]

\multicolumn{6}{c}{\textbf{Nuclear Magnetic Resonance (NMR)}}\\
\hline
$\sigma$-tensor    & 4c       & PT2 & \checkmark & -          & \citenum{Komorovsky2008,Komorovsky2010} \\
$J$-tensor         & 4c       & PT2 & \checkmark & -          & \citenum{Repisky2009} \\[8pt]

\multicolumn{6}{c}{\textbf{Paramagnetic Nuclear Magnetic Resonance (pNMR)}}\\
\hline
$\sigma$-tensor    & 4c       & PT2 & - & \checkmark & \citenum{Komorovsky2013,Jeremias2018} \\
$J$-tensor         & 4c       & PT2 & - & \checkmark & \citenum{Repisky2020} \\[8pt]

\multicolumn{6}{c}{\textbf{Optical Properties}}\\
\hline
UV/Vis EAS         & 1c,2c,4c & RT  & \checkmark & \checkmark & \citenum{Repisky2015,Konecny2016} \\
                   & 1c,2c,4c & DR  & \checkmark &            & \citenum{Konecny2019,Konecny2023} \\
                   & 1c,2c,4c & EV  & \checkmark & \checkmark & \citenum{Komorovsky2019,Konecny2023} \\[4pt]
X-ray EAS          & 1c,2c,4c & RT  & \checkmark & \checkmark & \citenum{Kadek2015} \\
                   & 1c,2c,4c & DR  & \checkmark &            & \citenum{Konecny2022,Konecny2023} \\
                   & 1c,2c,4c & EV  & \checkmark &            & \citenum{Konecny2023} \\[4pt] 
Polarizability     & 1c,2c,4c & RT  & \checkmark & \checkmark & \citenum{Konecny2016} \\
                   & 1c,2c,4c & DR  & \checkmark &            & \citenum{Konecny2019} \\
                   & 1c,2c,4c & EV  & \checkmark & \checkmark & \citenum{Komorovsky2019} \\[4pt]
Rad. lifetimes     & 1c,2c,4c & EV  & \checkmark & \checkmark & \citenum{Komorovsky2019,Repisky2020} \\[8pt]

\multicolumn{6}{c}{\textbf{Natural Chiroptical Properties}}\\
\hline
ECD                & 1c,2c,4c & RT  & \checkmark & \checkmark & \citenum{Konecny2018} \\
                   & 1c,2c,4c & DR  & \checkmark &            & \citenum{Konecny2019} \\[4pt]
ORD                & 1c,2c,4c & RT  & \checkmark & \checkmark & \citenum{Konecny2018} \\
                   & 1c,2c,4c & DR  & \checkmark &            & \citenum{Konecny2019} \\[8pt]

\multicolumn{6}{c}{\textbf{Optical Properties in Cavities}}\\
\hline
UV/Vis EAS         & 1c,2c,4c & EV  & \checkmark &            & \citenum{Konecny2024,Konecny2025} \\[8pt]

\multicolumn{6}{c}{\textbf{Time-Resolved Pump-Probe Spectroscopies}}\\
\hline
TR-EAS             & 1c,2c,4c & RT  & \checkmark & \checkmark & \citenum{Moitra2023} \\
TR-ECD             & 1c,2c,4c & RT  & \checkmark & \checkmark & \citenum{Moitra2025} \\[8pt]

\multicolumn{6}{c}{\textbf{Additional Properties}}\\
\hline
EFG                & 1c,2c,4c & PT1 & \checkmark &            & \citenum{Joosten2024}   \\
Mossbauer          & 1c,2c,4c & PT1 & \checkmark &            &    \\
NSR                & 1c,4c    & PT2 & \checkmark &            & \citenum{Malkin2013,Komorovsky2015-SR} \\[8pt]

\hline
\hline
\multicolumn{6}{l}{
\begin{minipage}{\textwidth}
\vspace{6pt}
\textbf{Abbreviations in alphabetical order:}
DR: damped response TDDFT, 
EFG: electric field gradient,
EV: eigenvalue TDDFT,
KR: Kramers-restricted,
KU: Kramers-unrestricted, 
NSR: nuclear spin-rotation constant,
ORD: optical rotatory dispersion,
PT1/2: static perturbation theory of the first/second order,
RT: real-time TDDFT,
(TR-)EAS: (time-resolved) electronic absorption spectroscopy,
(TR-)ECD: (time-resolved) electronic circular dichroism,
ZFS: zero-field splitting.
\vspace{6pt}
\end{minipage}
} \\
\hline

\end{longtable}
\renewcommand{\arraystretch}{1.0}

\section{\label{sec:theory}Theory and Exemplary Applications}

\subsection{\label{sec:theory-x2c}A: X2C Hartree--Fock and Kohn--Sham DFT}
A convenient starting point for discussing the X2C Hamiltonian models implemented in \ReSpect, within the mean-field Hartree–Fock or Kohn–Sham DFT framework, is to consider the four-component (4c) Fock equations, expressed for convenience in an orthonormal basis
\begin{eqnarray}
   \mat{F}^{\text{4c}} \mat{C}^{\text{4c}}
   =
   \mat{C}^{\text{4c}} \epsilon^{\text{4c}},
   \qquad
   \mat{F}^{\text{4c}} 
   = 
   \begin{pmatrix}
      \mat{F}_{11} & \mat{F}_{12} & \mat{F}_{13} & \mat{F}_{14} \\
      \mat{F}_{21} & \mat{F}_{22} & \mat{F}_{23} & \mat{F}_{24} \\
      \mat{F}_{31} & \mat{F}_{32} & \mat{F}_{33} & \mat{F}_{34} \\
      \mat{F}_{41} & \mat{F}_{42} & \mat{F}_{43} & \mat{F}_{44} \\
   \end{pmatrix}
   \in\mathbb{C}_{4n\times4n}
   .
\end{eqnarray}
Here, $\mat{F}^{\text{4c}}$ denotes the Fock matrix, while $\mat{C}^{\text{4c}}$ and
$\epsilon^{\text{4c}}$ are the corresponding eigenvector and eigenvalue matrices, respectively.
In the most general case, both $\mat{F}^{\text{4c}}$ and $\mat{C}^{\text{4c}}$ are full complex-valued matrices of size $4n\times4n$, where $n$ refers to the size of the user-selected scalar basis.
Compared to the common nonrelativistic one-component (1c) formalism, the 4c framework requires processing approximately 32 times more data and, in our experience, results in a 10-15-fold increase in computational cost~\cite{Hrda2014,Repisky2020}.

The central idea of the exact two-component (X2C) approach is to reduce the computational overhead by transforming the full Fock matrix into its block-diagonal form using a unitary decoupling matrix 
$\mat{U}\in\mathbb{C}^{4n\times4n}$~\cite{Heully1986,Jensen2005,Kutzelnigg2005,Liu2007,Ilias2007}: 
\begin{eqnarray}
      \mat{F}^{\text{4c}}
      \rightarrow
      \tilde{\mat{F}}^{\text{4c}}
      =
      \mat{U}^{\dagger}\mat{F}^{\text{4c}}\mat{U}
      =
      \begin{pmatrix}
         \tilde{\mat{F}}_{11} & \tilde{\mat{F}}_{12} & \mat{0}              & \mat{0} \\
         \tilde{\mat{F}}_{21} & \tilde{\mat{F}}_{22} & \mat{0}              & \mat{0} \\
         \mat{0}              & \mat{0}              & \tilde{\mat{F}}_{33} & \tilde{\mat{F}}_{34} \\
         \mat{0}              & \mat{0}              & \tilde{\mat{F}}_{43} & \tilde{\mat{F}}_{44} \\
      \end{pmatrix}
      .
\end{eqnarray}
Thanks to the unitary property of $\mat{U}$, all eigenvalues of the parent 4c problem can be reproduced to computer precision by solving two sets of uncoupled Fock equations, each with half the dimension of the original problem. By disregarding the set associated with negative-energy solutions, we are left with the X2C Fock equations (or matrix) for the positive-energy solutions (++):
\begin{eqnarray}
    \tilde{\mat{F}}^{\text{2c}} \tilde{\mat{C}}^{\text{2c}}
    =
    \tilde{\mat{C}}^{\text{2c}} \epsilon^{\text{2c}},
    \qquad
    \tilde{\mat{F}}^{\text{2c}}
    :=
    \big[ \tilde{\mat{F}}^{\text{4c}} \big]^{++}
    =
    \begin{pmatrix}
        \tilde{\mat{F}}_{11} & \tilde{\mat{F}}_{12} \\
        \tilde{\mat{F}}_{21} & \tilde{\mat{F}}_{22} \\
    \end{pmatrix}
    \in\mathbb{C}_{2n\times2n}
    .
\end{eqnarray}
Note that this block-diagonal structure of $\tilde{\mat{F}}^{\text{4c}}$ also carries over to the transformed solution matrix $\tilde{\mat{C}}^{\text{4c}}$, \emph{i.e.}
\begin{eqnarray}
      \tilde{\mat{C}}^{\text{4c}}
      =
      \mat{U}^{\dagger}\mat{C}^{\text{4c}}
      =
      \begin{pmatrix}
         \tilde{\mat{C}}_{11} & \tilde{\mat{C}}_{12} & \mat{0}              & \mat{0} \\
         \tilde{\mat{C}}_{21} & \tilde{\mat{C}}_{22} & \mat{0}              & \mat{0} \\
         \mat{0}              & \mat{0}                    & \tilde{\mat{C}}_{33} & \tilde{\mat{C}}_{34} \\
         \mat{0}              & \mat{0}                    & \tilde{\mat{C}}_{43} & \tilde{\mat{C}}_{44} \\
      \end{pmatrix}
      \quad\Rightarrow\quad
      \tilde{\mat{C}}^{\text{2c}}
      :=
      \big[ \tilde{\mat{C}}^{\text{4c}} \big]^{++}
      = 
      \begin{pmatrix}
         \tilde{\mat{C}}_{11} & \tilde{\mat{C}}_{12} \\
         \tilde{\mat{C}}_{21} & \tilde{\mat{C}}_{22} \\
      \end{pmatrix}
      \in\mathbb{C}_{2n\times2n}
      .
\end{eqnarray}
Before proceeding, let us recall that all two-component (2c) quantities undergoing an exact two-component (picture-change) transformation are marked with a tilde.

In practice, there are several flavors of X2C Hamiltonian models, differing in the choice of the 4c Hamiltonian used to construct the decoupling matrix $\mat{U}$—and thus the resulting two-component model. In \ReSpect, we have implemented the molecular mean-field X2C (mmfX2C) model, which is constructed \emph{a posteriori} from the converged molecular 4c self-consistent field (SCF) Fock matrix~\cite{Sikkema2009}. The mmfX2C model is widely used in connection with post-SCF electron correlation and/or response theory calculations, as its results exactly reproduce the positive-energy solutions of the original 4c Fock equations, making it an excellent reference method. However, constructing the mmfX2C model requires a full molecular 4c SCF calculation, which can be computationally demanding—particularly for large molecular systems. Therefore, it is common practice to look for solutions that carry out SCF iterations directly in 2c mode.
%In addition, the mmfX2C approach remains an approximation in any subsequent post-SCF calculation if picture-change untransformed two-electron integrals are used in place of their fully transformed counterparts.

The simplest, though rather inaccurate, approach with \emph{a priori} X2C decoupling uses only the one-electron (1e) terms of the Dirac Hamiltonian. The resulting X2C Hamiltonian model, referred to as the one-electron X2C (1eX2C) model, is therefore considered "exact" only with respect to the 1e terms, as the two-electron (2e) interactions are entirely omitted from the decoupling~\cite{Kutzelnigg2005,Liu2007,Ilias2007}. The subsequent SCF iterations are typically carried out using the Fock matrix,
\begin{eqnarray}
      \mat{\tilde{F}}^{\text{1eX2C}}
      =
      \mat{\tilde{h}}^{\text{2c}}
      +
      \mat{G}^{\text{2c}}[\mat{\tilde{D}}^{\text{2c}}]
      \qquad
      \begin{cases}
        \mat{\tilde{h}}^{\text{2c}}
        :=
        \Big[
        \mat{U}^{\dagger}
        \mat{h}^{\text{4c}}
        \mat{U}
        \Big]^{++}
        \\
        G^{\text{2c}}_{\mu\nu}[\mat{\tilde{D}}^{\text{2c}}]
        :=
        \sum_{\kappa\lambda}
        g^{\text{2c}}_{\mu\nu,\kappa\lambda}
        \tilde{D}^{\text{2c}}_{\lambda\kappa}
        \end{cases}
\end{eqnarray}
and involving the picture-change transformed one-electron term $\mat{\tilde{h}}^{\text{2c}}$ and the two-electron term $\mat{G}^{\text{2c}}$ with the standard, picture-change \emph{untransformed} two-electron integrals
\begin{eqnarray}
   g^{\text{2c}}_{\mu\nu,\kappa\lambda}
   =
   \mathcal{I}_{\mu\nu,\kappa\lambda}
   -
   \mathcal{I}_{\mu\lambda,\kappa\nu}
   ;\qquad
   \mathcal{I}_{\mu\nu,\kappa\lambda}
   :=
   \iint
   \chi^{\dagger}_{\mu}(\vec{r}_{1})
   \chi_{\nu}(\vec{r}_{1})
   r_{12}^{-1}
   \chi^{\dagger}_{\kappa}(\vec{r}_{2})
   \chi_{\lambda}(\vec{r}_{2})   
   d^{3}\vec{r}_{1}d^{3}\vec{r}_{2}
   .
\end{eqnarray}
As demonstrated below, the 1eX2C Hamiltonian model represents a rather severe approximation, introducing the so-called two-electron picture-change error (2ePCE)~\cite{Knecht2022}. The extension of 1eX2C to Kohn--Sham DFT is straightforward and has also been implemented in \ReSpect. However, the absence of a picture-change transformed exchange–correlation (xc) contribution introduces an additional error, known as the xc picture-change error (xcPCE)~\cite{Knecht2022}.

To overcome previous limitations, two in-house X2C Hamiltonian models were developed and implemented in \ReSpect{} in collaboration with the DIRAC team: the atomic mean-field exact two-component (amfX2C) and the extended amfX2C (eamfX2C) approaches~\cite{Knecht2022}.
These models obviate the need for a full 4c SCF reference and offer a simple, computationally efficient, and numerically accurate way to eliminate two-electron and exchange–correlation picture-change errors. A key observation here is that the exact two-component Fock matrix,
\begin{eqnarray}
   \mat{\tilde{F}}^{\text{2c}}
   =
   \mat{\tilde{h}}^{\text{2c}}
   +
   \mat{\tilde{G}}^{\text{2c}}[\mat{\tilde{D}}^{\text{2c}}]
   ,\qquad
   \tilde{G}^{\text{2c}}_{\mu\nu}[\mat{\tilde{D}}^{\text{2c}}]
   =
   \sum_{\kappa\lambda}
   \tilde{g}^{\text{2c}}_{\mu\nu,\kappa\lambda}
   \tilde{D}^{\text{2c}}_{\lambda\kappa}
   ,
\end{eqnarray}
requires the picture-change transformed density matrix ($\tilde{\mat{D}}^{\text{2c}}$), as well as the one- and two-electron integrals, denoted $\tilde{\mat{h}}^{\text{2c}}$ and $\tilde{\mat{g}}^{\text{2c}}$, respectively:
\begin{align}
      \mat{\tilde{D}}^{\text{2c}}
      & :=
      \Big[
      \mat{U}^{\dagger}
      \mat{D}^{\text{4c}}
      \mat{U}
      \Big]^{++}
      \nonumber
      \\
      \mat{\tilde{h}}^{\text{2c}}
      & :=
      \Big[
      \mat{U}^{\dagger}
      \mat{h}^{\text{4c}}
      \mat{U}
      \Big]^{++}
      \\
      \mat{\tilde{g}}^{\text{2c}}
      & :=
      %\underbrace{
      \Big[
      \mat{U}^{\dagger}
      \mat{U}^{\dagger}
      \mat{g}^{\text{4c}}
      \mat{U}
      \mat{U}
      \Big]^{++}.
      \nonumber
\end{align}
Although the evaluation of $\tilde{\mat{h}}^{\text{2c}}$ and $\tilde{\mat{D}}^{\text{2c}}$ is relatively inexpensive due to their reliance on two-index transformations, the PC transformation of 4c two-electron integrals $\mat{g}^{\text{4c}}$ involves a costly four-index transformation, making thus 2c calculations more demanding than their 4c counterparts. 
%Therefore, it is common practice to neglect the two-electron PC correction entirely and construct the Fock matrix using \emph{untransformed} two-electron integrals $\mat{g}^{\text{2c}}$, \emph{i.e.}
%\begin{eqnarray}
   %\tilde{g}^{\text{2c}}_{\mu\nu,\kappa\lambda}
   %\simeq
   %g^{\text{2c}}_{\mu\nu,\kappa\lambda}
   %=
   %\mathcal{I}_{\mu\nu,\kappa\lambda}
   %-
   %\mathcal{I}_{\mu\lambda,\kappa\nu}
   %;\qquad
   %\mathcal{I}_{\mu\nu,\kappa\lambda}
   %:=
   %\iint
   %\chi^{\dagger}_{\mu}(\vec{r}_{1})
   %\chi_{\nu}(\vec{r}_{1})
   %r_{12}^{-1}
   %\chi^{\dagger}_{\kappa}(\vec{r}_{2})
   %\chi_{\lambda}(\vec{r}_{2})   
   %d^{3}\vec{r}_{1}d^{3}\vec{r}_{2}
   %.
%\end{eqnarray}
%As discussed above, this is a rather severe approximation that introduces the so-called two-electron picture-change error (2ePCE). 

The central idea of the amfX2C Hamiltonian model is to construct $\tilde{\mat{G}}^{\text{2c}}$ as the sum of the two-electron contribution with untransformed two-electron integrals ($\mat{G}^{\text{2c}}$) and a picture-change correction complement ($\Delta\mat{G}^{\text{2c}}$):
\begin{eqnarray}
      \label{eq:G-in-amfX2C}
      \mat{\tilde{G}}^{\text{2c}}[\mat{\tilde{D}}^{\text{2c}}] 
      = 
      \mat{G}^{\text{2c}}[\mat{\tilde{D}}^{\text{2c}}]
      + 
      \Delta\mat{G}^{\text{2c}}[\mat{\tilde{D}}^{\text{2c}}]
      ,\qquad
      G^{\text{2c}}_{\mu\nu}[\mat{\tilde{D}}^{\text{2c}}]
      =
      \sum_{\kappa\lambda}
      g^{\text{2c}}_{\mu\nu,\kappa\lambda}
      \tilde{D}^{\text{2c}}_{\lambda\kappa}
      .
\end{eqnarray}
Hence, the Fock matrix reads
\begin{eqnarray}
      \mat{\tilde{F}}^{\text{2c}}
      =
      \mat{\tilde{h}}^{\text{2c}}
      +
      \mat{G}^{\text{2c}}[\mat{\tilde{D}}^{\text{2c}}]
      +
      \Delta\mat{G}^{\text{2c}}[\mat{\tilde{D}}^{\text{2c}}]
      .
\end{eqnarray}
Numerical analysis reveals that the correction term exhibits a localized atomic character and can therefore be accurately approximated as a superposition of atomic contributions
\begin{eqnarray}
      \Delta\mat{G}^{\text{2c}}
      \simeq
      \Delta\mat{G}^{\text{2c}}_{\bigoplus}
      %=
      %\bigoplus_{K=1}^{\text{atoms}}
      %\Delta\mat{G}^{\text{2c}}_{K}[\mat{\tilde{D}}^{\text{2c}}_{K}]
      :=
      \bigoplus_{K=1}^{\text{atoms}}
      \mat{\tilde{G}}^{\text{2c}}_{K}[\mat{\tilde{D}}^{\text{2c}}_{K}] 
      - 
      \mat{G}^{\text{2c}}_{K}[\mat{\tilde{D}}^{\text{2c}}_{K}]
      .
\end{eqnarray}
Here, the subscript $K$ runs over all atoms in a molecular system and labels the atomic building blocks of the 2ePC corrections. This approach defines our amfX2C Hamiltonian model for the Hartree--Fock method
\begin{eqnarray}
      \mat{\tilde{F}}^{\text{amfX2C}}
      =
      \mat{\tilde{h}}^{\text{2c}}
      +
      \mat{G}^{\text{2c}}[\mat{\tilde{D}}^{\text{2c}}]
      +
      \Delta\mat{G}^{\text{2c}}_{\bigoplus}
      ,
\end{eqnarray}
where $\Delta\mat{G}^{\text{2c}}_{\bigoplus}$ is a static Fock contribution evaluated prior to the molecular 2c SCF procedure by assembling the results of 4c atomic Kramers-restricted fractional-occupation calculations, as discussed in Ref.~\citenum{Knecht2022}. In contrast, $\mat{G}^{\text{2c}}[\tilde{\mat{D}}^{\text{2c}}]$ is a dynamic term that is updated throughout the SCF iterations and requires only conventional nonrelativistic two-electron integrals. Based on our experience, the computational cost of amfX2C SCF calculations does not exceed a factor of four compared to nonrelativistic 1c calculations, while achieving energy accuracy within approximately 10 $\mu$Hartree per atom relative to the full 4c Hamiltonian~\cite{Knecht2022,Konecny2023}. 

Importantly, the amfX2C Hamiltonian model has a correct atomic limit as it reproduces atomic 4c SCF calculations exactly at the 2c level. The model incorporates both scalar-relativistic and spin-orbit two-electron picture-change corrections on an equal footing, with their evaluation scaling linearly with system size. The algebraic nature of amfX2C also allows an easy extraction of 2ePC corrections not only from the common 2e Coulomb Hamiltonian but also from more elaborate Gaunt and Breit Hamiltonians. The amfX2C model can also be extended to Kohn–Sham DFT, where it takes the form~\cite{Knecht2022}:
\begin{eqnarray}
      \mat{\tilde{F}}^{\text{amfX2C}}
      =
      \mat{\tilde{h}}^{\text{2c}}
      +
      \mat{G}^{\text{2c}}\big[\mat{\tilde{D}}^{\text{2c}}\big]
      +
      \Delta\mat{G}^{\text{2c}}_{\bigoplus}
      +
      \mat{V}^{\text{2c}}_{\text{xc}}\big[\vec\rho^{\text{2c}}\big]
      +
      \Delta\mat{V}^{\text{2c,xc}}_{\bigoplus}
      .
\end{eqnarray}
Here, the last two terms arise from the exact PC-transformed exchange--correlation (xc) interaction term 
%$\tilde{\mat{V}}^{\text{2c}}_{\text{xc}}\big[\tilde{\vec{\rho}}^{\text{2c}}\big]$, 
\begin{eqnarray}
      \mat{\tilde{V}}^{\text{2c}}_{\text{xc}}\big[\tilde{\vec{\rho}}^{\text{2c}}\big]
      :=
      \int
      v^{\text{xc}}_k[\tilde{\vec{\rho}}^{\text{2c}}](\vec{r})
      \mat{\tilde{\Omega}}_k^{\text{2c}}(\vec{r}) d^{3}\vec{r}
      \qquad
      \begin{cases}
         \mat{\tilde{\Omega}}_k^{\text{2c}}
         :=
         \Big[
         \mat{U}^{\dagger}
         \mat{\Omega}_k^{\text{4c}}
         \mat{U}
         \Big]^{++}
         \\[0.2cm]     
         \tilde{\rho}_k^{\text{2c}}
         :=
         \Tr
         \Big\{
         \mat{\tilde{\Omega}}_k^{\text{2c}}
         \mat{\tilde{D}}^{\text{2c}}
         \Big\}
         \\[0.2cm]
         v^{\text{xc}}_k[\tilde{\vec{\rho}}^{\text{2c}}] := \frac{\partial \varepsilon^{\text{xc}}}{\partial \tilde{\rho}_k^{\text{2c}}}-\left( \vec{\nabla}\cdot\frac{\partial\varepsilon^{\text{xc}}}{\vec{\nabla}\tilde{\rho}_k^{\text{2c}}}\right)
         \\[0.2cm]
         \Omega^{\text{4c}}_{k,\mu\nu}
         :=
         (X^{\text{4c}}_{\mu})^{\dagger} \Sigma_k
         X^{\text{4c}}_{\nu}
      \end{cases}
\end{eqnarray}
Here $\varepsilon^{\text{xc}}$ is the exchange-correlation energy density, $\vec{X}^{\text{4c}}$ is the restricted kinetic balance basis (RKB) [see Eq.~\eqref{eq:RMB} with $\vec{A}=0$], and $\Sigma_k$ are 4c identity $(k\!=\!0)$ and spin operators $(k\!=\!1,2,3)$. The exchange-correlation potential $v_k^{\text{xc}}$ is a functional of the electronic charge $(k\!=\!0)$ and spin $(k\!=\!1,2,3)$ densities $\tilde\rho^{\text{2c}}_k(\vec{r})$, and uses the so-called noncollinear ansatz that is crucial for proper distribution of spin density in relativistic theory. For details on the noncollinear formulation of $v_k^{\text{xc}}$ in \ReSpect{} program interested readers are referred to Refs.~\citenum{Komorovsky2019,Repisky2020}.
The point-wise picture-change transformation of the overlap distribution matrix $\mat{\tilde{\Omega}}^{\text{2c}}$, even when using local approximations, introduces significant computational cost. Therefore, similar to the treatment of $\mat{\tilde{G}}^{\text{2c}}$ in Eq.~\eqref{eq:G-in-amfX2C}, the amfX2C model constructs $\mat{\tilde{V}}^{\text{2c}}_{\text{xc}}\big[\tilde{\vec{\rho}}^{\text{2c}}\big]$ from its untransformed counterpart $\mat{V}^{\text{2c}}_{\text{xc}}\big[\vec{\rho}^{\text{2c}}\big]$, along with an additive picture-change correction term $\Delta\mat{V}^{\text{2c}}_{\text{xc}}\big[\tilde{\vec{\rho}}^{\text{2c}}\big]$:
\begin{eqnarray}
      \mat{\tilde{V}}^{\text{2c}}_{\text{xc}}\big[\tilde{\vec{\rho}}^{\text{2c}}\big]
      =
      \mat{V}^{\text{2c}}_{\text{xc}}\big[\vec{\rho}^{\text{2c}}\big]
      +
      \Delta\mat{V}^{\text{2c}}_{\text{xc}}\big[\tilde{\vec{\rho}}^{\text{2c}}\big]
      ,\qquad
      \mat{V}^{\text{2c}}_{\text{xc}}\big[\vec{\rho}^{\text{2c}}\big]
      :=
      \int
      v^{\text{LDA}}_{\text{xc}}[\vec{\rho}^{\text{2c}}](\vec{r})
      \mat{\Omega}^{\text{2c}}(\vec{r}) d^{3}\vec{r}
      .
\end{eqnarray} 
The amfX2C picture-change correction to the xc potential is then obtained from atomic quantities as
\begin{eqnarray}
      \Delta\mat{V}^{\text{2c}}_{\text{xc}}
      \simeq
      \Delta\mat{V}^{\text{2c,xc}}_{\bigoplus}
      %=
      %\bigoplus_{K=1}^{\text{atoms}}
      %\Delta\mat{V}^{\text{2c}}_{\text{xc},K}
      :=
      \bigoplus_{K=1}^{\text{atoms}}
      \mat{\tilde{V}}^{\text{2c}}_{\text{xc},K}\big[\tilde{\vec{\rho}}^{\text{2c}}_{K}\big]
      -
      \mat{V}^{\text{2c}}_{\text{xc},K}\big[\vec{\rho}^{\text{2c}}_{K}\big]
      .
\end{eqnarray}

The main computational advantage of the amfX2C approach is that its two-electron and exchange-correlation picture-change (PC) corrections $\Delta{\mat{G}}^{\text{2c}}_{\bigoplus}$ and $\Delta{\mat{V}}^{\text{2c,xc}}_{\bigoplus}$ are constructed from atomic components. Consequently, all matrix elements $\Delta{G}^{\text{2c}}_{\bigoplus,\mu\nu}$ and 
$\Delta{V}^{\text{2c,xc}}_{\bigoplus,\mu\nu}$ are strictly zero when the basis functions $\mu$ and $\nu$ belong to different atomic centers. To account for inter-atomic PC corrections, we have designed and implemented the extended amfX2C Hamiltonian model (eamfX2C), which retains the same philosophy and form as the original amfX2C model but differs in the PC correction terms~\cite{Knecht2022}
\begin{align}
      \text{HF:}~~~
      \mat{\tilde{F}}^{\text{eamfX2C}}
      & =
      \mat{\tilde{h}}^{\text{2c}}
      +
      \mat{G}^{\text{2c}}[\mat{\tilde{D}}^{\text{2c}}]
      +
      \Delta\mat{G}^{\text{2c}}[\mat{\tilde{D}}^{\text{2c}}_{\bigoplus}]
      \\
      \text{KS:}~~~
      \mat{\tilde{F}}^{\text{eamfX2C}}
      & =
      \mat{\tilde{h}}^{\text{2c}}
      +
      \mat{G}^{\text{2c}}[\mat{\tilde{D}}^{\text{2c}}]
      +
      \Delta\mat{G}^{\text{2c}}[\mat{\tilde{D}}^{\text{2c}}_{\bigoplus}]
      +
      \mat{V}^{\text{2c}}_{\text{xc}}[\vec\rho^{\text{2c}}]
      +
      \Delta\mat{V}^{\text{2c}}_{\text{xc}}[\tilde{\vec\rho}^{\text{2c}}_{\bigoplus}]
      .
\end{align}
Here, $\Delta\mat{G}^{\text{2c}}$ and $\Delta\mat{V}^{\text{2c}}_{\text{xc}}$ are evaluated in a \emph{full} molecular basis using an approximate density matrix $\mat{\tilde{D}}^{\text{2c}}_{\bigoplus}$ (or a charge density $\tilde{\vec\rho}^{\text{2c}}_{\bigoplus}$) obtained by superposing the corresponding atomic terms:
\begin{align}
      \Delta\mat{G}^{\text{2c}}[\mat{\tilde{D}}^{\text{2c}}]
      & \simeq
      \Delta\mat{G}^{\text{2c}}[\mat{\tilde{D}}^{\text{2c}}_{\bigoplus}]
      =
      \mat{\tilde{G}}^{\text{2c}}[\mat{\tilde{D}}^{\text{2c}}_{\bigoplus}] 
      - 
      \mat{G}^{\text{2c}}[\mat{\tilde{D}}^{\text{2c}}_{\bigoplus}]
      \\
      \Delta\mat{V}^{\text{2c}}_{\text{xc}}\big[\tilde{\vec\rho}^{\text{2c}}\big]
      & \simeq
      \Delta\mat{V}^{\text{2c}}_{\text{xc}}\big[\tilde{\vec\rho}^{\text{2c}}_{\bigoplus}\big]
      =
      \mat{\tilde{V}}^{\text{2c}}_{\text{xc}}\big[\tilde{\vec\rho}^{\text{2c}}_{\bigoplus}\big]
      -
      \mat{V}^{\text{2c}}_{\text{xc}}\big[\vec\rho^{\text{2c}}_{\bigoplus}\big]
      .
\end{align}
For all details regarding to (e)amfX2C models, interested readers are referred to Ref.~\citenum{Knecht2022}.

To showcase the numerical accuracy of our implemented X2C Hamiltonian models, we performed HF and KS-DFT calculations on gold dimer with bond-distance 4.67 atomic units, making use of Dyall's atom-centered uncontracted Gaussian-type basis sets of valence triple-$\zeta$ quality (dyall-vtz)~\cite{Dyall2010-5d}, a hybrid PBE0 exchange--correlation functional~\cite{Perdew1996,Perdew1997Erratum,Adamo1999,Slater1951}, a point nucleus model, and an explicit inclusion of $(\text{SS}|\text{SS})$-type electron-repulsion AO integrals to ease comparison. 

\begin{table}[]
\centering
\begin{tabular}{l r r r r }
\hline \hline
 & 1eX2C & amfX2C & eamfX2C & 4C \\
\hline
$E$ & -380\highlight{65.91336}    & -38079.021\highlight{63} & -38079.0217\highlight{4} & -38079.02175 \\
... \\
$\epsilon_{1s_{1/2}}$ & -298\highlight{3.57361} & -2987.714\highlight{36} & -2987.714\highlight{37} & -2987.71442 \\
$\epsilon_{2s_{1/2}}$ & -53\highlight{1.80960}  & -532.4225\highlight{8}  & -532.42259              & -532.42259 \\
$\epsilon_{2p_{1/2}}$ & -5\highlight{10.48203}  & -509.2815\highlight{8}  & -509.2815\highlight{9}  & -509.28157 \\
$\epsilon_{2p_{3/2}}$ & -44\highlight{0.76539}  & -441.6930\highlight{2}  & -441.6930\highlight{4}  & -441.69305 \\
                      & -44\highlight{0.76500}  & -441.69265              & -441.69265              & -441.69265 \\
... \\
$\epsilon_{75}$       & -0.41\highlight{771} & -0.41821             & -0.41821             & -0.41821 \\
$\epsilon_{76}$       & -0.4\highlight{0742} & -0.411\highlight{68} & -0.41170             & -0.41170 \\
$\epsilon_{77}$       & -0.3\highlight{8817} & -0.3919\highlight{6} & -0.39198             & -0.39198 \\
$\epsilon_{78}$       & -0.3\highlight{8783} & -0.39039             & -0.39039             & -0.39039 \\
$\epsilon_{79}$       & -0.295\highlight{69} & -0.2958\highlight{9} & -0.29588             & -0.29588 \\
\hline \hline
\end{tabular}
\caption{SCF total energy $E$ and spinor energies in atomic units for selected doubly occupied spinors ($\epsilon$) of Au$_2$ as obtained from HF/dyall-vtz calculations with all discussed X2C Hamiltonian models in comparison with reference four-component (4C) Dirac-Coulomb results. Note that mmfX2C results are identical with 4C and the differences are marked in red.}
\label{tab:hf-scf}
\end{table}

\begin{table}[]
\centering
\begin{tabular}{l r r r r }
\hline \hline
 & 1eX2C & amfX2C & eamfX2C & 4C \\
\hline
$E$ & -380\highlight{77.24610} & -38093.827\highlight{36} & -38093.827\highlight{50} & -38093.82740 \\
... \\
$\epsilon_{1s_{1/2}}$ & -296\highlight{4.34387} & -2969.876\highlight{19} & -2969.8762\highlight{0} & -2969.87626 \\
$\epsilon_{2s_{1/2}}$ & -52\highlight{4.25485}  & -525.005\highlight{69}  & -525.00570              & -525.00570 \\
$\epsilon_{2p_{1/2}}$ & -503.\highlight{70410}  & -503.0808\highlight{2}  & -503.0808\highlight{2}  & -503.08080 \\
$\epsilon_{2p_{3/2}}$ & -434.5\highlight{4808}  & -435.583\highlight{59}  & -435.5836\highlight{0}  & -435.58361 \\
                      & -434.5\highlight{4798}  & -435.5834\highlight{9}  & -435.58348              & -435.58348 \\
... \\
$\epsilon_{75}$       & -0.2\highlight{8745} & -0.2906\highlight{0} & -0.29062             & -0.29062 \\
$\epsilon_{76}$       & -0.27\highlight{232} & -0.2756\highlight{6} & -0.2756\highlight{8} & -0.27569 \\
$\epsilon_{77}$       & -0.2\highlight{6880} & -0.2704\highlight{9} & -0.27048             & -0.27048 \\
$\epsilon_{78}$       & -0.26\highlight{037} & -0.263\highlight{58} & -0.26360             & -0.26360 \\
$\epsilon_{79}$       & -0.2\highlight{5956} & -0.2605\highlight{8} & -0.26057             & -0.26057 \\
\hline \hline
\end{tabular}
\caption{SCF total energy $E$ and spinor energies in atomic units for selected doubly occupied spinors ($\epsilon$) of Au$_2$ as obtained from DFT/PBE0/dyall-vtz calculations with all discussed X2C Hamiltonian models in comparison with reference four-component (4C) Dirac-Coulomb results. Note that mmfX2C results are identical with 4C and the differences are marked in red.}
\label{tab:pbe0-scf}
\end{table}

Tables~\ref{tab:hf-scf} and~\ref{tab:pbe0-scf} present the total energy ($E$) and selected occupied spinor energies ($\epsilon$)--spanning selected core to valence levels, computed at the HF and KS-DFT levels, respectively. The deviations from reference 4C Dirac–Coulomb data are highlighted in red. As previously discussed, different X2C Hamiltonians vary in their treatment of 2ePC corrections. The most significant discrepancies between the X2C and 4C frameworks occur for the innermost $s$ and $p$ shells, where 2ePCE corrections are critical.
For example, the 1eX2C model, which neglects 2ePCE entirely, gives highest deviation for $\epsilon_{1s_{1/2}}$ of +4.14 Hartree (HF) and +5.53 Hartree (PBE0). In contrast, the amfX2C and eamfX2C schemes, achieve excellent agreement with the 4C reference, with maximum discrepancies of the order of $10^{-4}$ Hartree. These results underscore the numerical accuracy of the amf-based 2ePCE corrections, particularly for core regions near the nucleus. This is further demonstrated in the subsequent sections focused on core-level spectroscopies.

%%%%%%%%%%%%%%%%%%%%%%%%%%%%%%%%%%%%%%%%%%%%%%%%%%%%%%%%%%%%%%%%%%%%
\subsection{\label{sec:applications-EPR}B: X2C Electron Paramagnetic Resonance}

Let us begin our discussion of EPR parameters evaluation by considering the electronic energy of a molecular system subjected to a static magnetic field. In this context, the canonical momentum of an electron $\vec{p}$ is replaced by the mechanical momentum $\vec{p} + \frac{1}{c}\vec{A}$, where $\vec{A}$ is the vector potential associated with the magnetic field. Throughout this section, we use the Hartree system of atomic units, also known as Gauss-based atomic units.

The electronic energy of the coupled molecule-field system can be written within 4c KS-DFT as a trace of the product of the density matrix with the one-electron $\mat{h}^{\text{4c}}$, two-electron $\mat{G}^{\text{4c}}$, and the exchange-correlation part $E_{\text{xc}}$:
\begin{eqnarray}
   E_{v}
   =
   \Tr\big\{\mat{h}^{\text{4c}}\mat{D}^{\text{4c}}_{v}\big\}
   +
   \frac{1}{2}
   \Tr\big\{\mat{G}^{\text{4c}}\mat{D}^{\text{4c}}_{v}\big\}
   +
   E_{\text{xc}}\big[\vec{\rho}^{\text{4c}}_{v}\big]
   \qquad
   \begin{cases}
      \mat{h}^{\text{4c}} = \mat{h}^{\text{4c}}[\vec{X}^{\text{4c}}(\vec{A}),\vec{A}]   
      \\
      \mat{G}^{\text{4c}} = \mat{G}^{\text{4c}}[\vec{X}^{\text{4c}}(\vec{A})]   
      \\
      \vec{\rho}^{\text{4c}}_{v} = \vec{\rho}^{\text{4c}}_{v}[\vec{X}^{\text{4c}}(\vec{A}),\mat{D}^{\text{4c}}_{v}(\vec{A})]
   \end{cases}
\end{eqnarray}
Here, $E_{v}:=E(\vec{J}_v)$ and $\mat{D}^{\text{4c}}_{v}:=\mat{D}^{\text{4c}}(\vec{J}_v)$, where $\vec{J}_v$ ($v=1,2,3$) is the magnetization vector of the $v$th Kohn–Sham (KS) determinant. Each of the three magnetization vectors corresponds to a KS determinant obtained from an independent SCF calculation. A comprehensive discussion of the theoretical framework for calculating EPR parameters within DFT can be found in Ref.~\citenum{Komorovsky2024, Komorovsky2023}. It is important to note that the vector potential enters the energy expression both explicitly, through the one-electron term, and implicitly through the 4c basis functions and density matrix.

The dependence of the 4c basis on the vector potential arises directly from the application of minimal electromagnetic coupling to the RKB operator, which generates the small-component basis functions. This procedure gives rise to the restricted magnetic balance (RMB) basis~\cite{Komorovsky2008, Repisky2009, Komorovsky2010}
\begin{align}
      \label{eq:RMB}
      X_{\mu}^{\text{4c,RMB}}(\vec{r},\vec{A})
      & =
      \begin{bmatrix}
         X_{\mu}^{\text{L}}(\vec{r}) & 0_{2}
         \\
         0_{2} & ~\qquad X_{\mu}^{\text{S,RMB}}(\vec{r},\vec{A})
      \end{bmatrix}
      =
      \begin{bmatrix}
         \text{I}_{2} & 0_{2}
         \\
         0_{2} & \frac{1}{2c}\vec{\sigma}\cdot\big(\vec{p}+\frac{1}{c}\vec{A}\big)
      \end{bmatrix}
      \chi_{\mu}(\vec{r}).
\end{align}
The concept of RMB basis was introduced into relativistic quantum chemistry approximately 15 years ago as a natural extension of the RKB basis in cases involving magnetic fields. In addition, to ensure gauge invariance of the g-tensor and rapid convergence of the results with respect to basis set size, the RMB basis is complemented by London's phase factor. This factor shifts the gauge origin of the vector potential from an arbitrary point $\vec{R}_0$ to the center of each scalar basis function, $\chi_\mu$:
\begin{align}
      X_{\mu}^{\text{4c,RMB--GIAO}}(\vec{r},\vec{A})
      & =
      \begin{bmatrix}
         X_{\mu}^{\text{L,GIAO}}(\vec{r}) & 0_{2}
         \\
         0_{2} & X_{\mu}^{\text{S,RMB--GIAO}}(\vec{r},\vec{A})
      \end{bmatrix}
      \nonumber
      \\
      & =
      \begin{bmatrix}
         \text{I}_{2} & 0_{2}
         \\
         0_{2} & \frac{1}{2c}\vec{\sigma}\cdot\big(\vec{p}+\frac{1}{c}\vec{A}\big)
      \end{bmatrix}
      \exp\left\{\frac{i}{c}(\vec{A}_{0\mu}\cdot\vec{r})\right\}
      \chi_{\mu}(\vec{r}),
\end{align}
where $\vec{A}_{0\mu}=\frac{1}{2}[\vec{B}\times(\vec{R}_0-\vec{R}_\mu)]$.

Now, let us take a closer look at the EPR g-tensor, which parametrizes the interaction energy of the total electronic magnetic moment of a system and a static uniform magnetic field $\vec{B}$ originating from an external source. In the Coulomb gauge, the corresponding vector potential depends on an arbitrary gauge origin $\vec{R}_0$, and is given by:
\begin{eqnarray}
      \vec{A}(\vec{r})
      :=
      \vec{A}_{0}(\vec{r})
      = 
      \frac{1}{2}(\vec{B}\times\vec{r}_{0})
      ,\qquad
      \vec{r}_{0} = \vec{r} - \vec{R}_{0}.
\end{eqnarray}
By differentiating the 4c Lagrange functional---comprising the electronic energy for a given
magnetization $\vec{J}_v$ and the orthonormality constraints imposed on the MOs---with respect to the components of the magnetic field, one arrives at the final 4c expression for the g-tensor
\begin{align}
      g_{uv}
      &=
      \frac{2c}{S}
      \left[
      \frac{\partial L_v}{\partial B_u} +
      \frac{\partial L_v}{\partial {X_{\mu}          }} \frac{d {X_{\mu}          }}{d B_{u}} +
      \frac{\partial L_v}{\partial {X_{\mu}^{\dagger}}} \frac{d {X_{\mu}^{\dagger}}}{d B_{u}}
      \right]\Bigg|_{\vec{B}=0}
      \nonumber
      \\
      & =
      \frac{2c}{S}
      \bigg[
      \Tr\big\{
        \mat{h}^{\text{4c}[B_{u}]}
        \mat{D}^{\text{4c}[0]}_{v}
      \big\}
      -
      \varepsilon_{v,i}^{4c[0]}
      \Tr\big\{
        \mat{S}^{\text{4c}[B_{u}]}
        \mat{d}^{\text{4c}[0]}_{v,i}
      \big\}
      \nonumber
      \\
      \label{eq:gt-4c}
      & +
      \Tr\big\{
         \mat{G}^{\text{4c}[B_{u}]}
         \mat{D}^{\text{4c}[0]}_{v}
      \big\}
      +
      \Tr\big\{
        \mat{V}^{\text{4c}[B_{u}]}_{\text{xc}}
        \mat{D}^{\text{4c}[0]}_{v}
      \big\}
      \bigg],
\end{align}
where $\varepsilon_{v,i}^{4c[0]}$ are occupied one-electron energies obtained from the 4c SCF procedure with magnetization $\vec{J}_v$. In addition, $\mat{S}^{\text{4c}}$ is the 4c overlap matrix and $\mat{d}^{\text{4c}[0]}_{v,i}$ is the density matrix of $i$th molecular orbital. Note, that the linear response matrices $\mat{S}^{\text{4c}[B_{u}]}$, $\mat{G}^{\text{4c}[B_{u}]}$, and $\mat{V}^{\text{4c}[B_{u}]}_{\text{xc}}$ depend on the perturbation parameter $B_u$ only through the RMB--GIAO basis. In comparison, the matrix $\mat{h}^{\text{4c}[B_{u}]}$ has both an explicit and implicit dependence on the perturbation parameter $B_u$. As we can see, the expression in Eq.~\eqref{eq:gt-4c} involves the trace of the unperturbed zero-order density matrix with one-electron, two-electron, and xc integrals, all differentiated to first order with respect to the magnetic field. Given that the 4c density matrix can be expressed in terms of the 2c density matrix and the X2C decoupling matrix $\mat{U}$, $\mat{D}^{\text{4c}[0]}_{v}=\mat{U}\mat{\tilde{D}}^{\text{2c}[0]}_{v}\mat{U}^{\dagger}$, $\mat{d}^{\text{4c}[0]}_{v,i}=\mat{U}\mat{\tilde{d}}^{\text{2c}[0]}_{v,i}\mat{U}^{\dagger}$ one can derive the final amfX2C expression for the g-tensor in RMB--GIAO basis
\begin{align}
      g_{uv} 
      & = 
      \frac{2c}{S} 
      \bigg[
      \Tr\big\{
        \mat{\tilde{h}}^{\text{2c}[B_{u}]} 
        \mat{\tilde{D}}^{\text{2c}[0]}_{v}
      \big\}
      -
      \varepsilon_{v,i}^{2c[0]}
      \Tr\big\{
        \mat{\tilde{S}}^{\text{2c}[B_{u}]}
        \mat{\tilde{d}}^{\text{2c}[0]}_{v,i}
      \big\}
      \nonumber
      \\
      & +
      \Tr\big\{
        \mat{\tilde{G}}^{\text{2c}[B_{u}]} 
        \mat{\tilde{D}}^{\text{2c}[0]}_{v}
      \big\}
      +
      \Tr\big\{
        \mat{\tilde{V}}^{\text{2c}[B_{u}]}_{\text{xc}} 
        \mat{\tilde{D}}^{\text{2c}[0]}_{v}
      \big\}
      \bigg]
      \nonumber
      \\
      &
      \begin{cases}
         \mat{\tilde{h}}^{\text{2c}[B_{u}]}
         =
         \Big[
         \mat{U}^{\dagger}
         \mat{h}^{\text{4c}[B_{u}]}
         \mat{U}
         \Big]^{++}
         \\
         \mat{\tilde{S}}^{\text{2c}[B_{u}]}
         =
         \Big[
         \mat{U}^{\dagger}
         \mat{S}^{\text{4c}[B_{u}]}
         \mat{U}
         \Big]^{++}
         \\
         \mat{\tilde{V}}^{\text{2c}[B_{u}]}_{\text{xc}} 
         =
         \Big[
         \mat{U}^{\dagger}
         \mat{V}^{\text{4c}[B_{u}]}_{\text{xc}} 
         \mat{U}
         \Big]^{++}
         \\
         \mat{\tilde{G}}^{\text{2c}[B_{u}]}
         =
         \Big[
         \mat{U}^{\dagger}
         \mat{G}^{\text{4c}[B_{u}]}
         \mat{U}
         \Big]^{++}
      \end{cases}
\end{align}
Similar to the SCF case, one must apply the picture-change transformation to the one-electron, two-electron, and xc contributions. For a detailed discussion of the two-component theory for calculating the g-tensor using London atomic orbitals and the RMB basis, the interested reader is referred to Refs.~\citenum{Repisky2025,MisenkovaPhD}.

The same logic also applies to the derivation of the working equations for the EPR hyperfine coupling A-tensor. This tensor parametrizes the interaction energy between the nuclear magnetic dipole moment and the internal magnetic field generated by the moving electrons. Here, the electromagnetic vector potential is associated with the static point magnetic dipole moments of the nucleus $N$,
\begin{eqnarray}
      \vec{A}(\vec{r})
      : =
      \sum_{N}
      \vec{A}_{N}(\vec{r})
      ,\qquad
      \vec{A}_{N}(\vec{r}) = \frac{ \gamma^{N}(\vec{I}^{N}\times\vec{r}_{N}) } {r_{N}^{3}}
      ,\qquad 
      \vec{r}_{N} = \vec{r} - \vec{R}_{N}.
\end{eqnarray}
By differentiating the 4c Lagrange functional of a particular magnetization $\vec{J}_v$ with respect to the components of nuclear spin $\vec{I}^N$, one gets the final 4c expression for the A-tensor
\begin{align}
      A_{uv}^{N}
      &=
      \frac{\gamma^{N}}{S}
      \left[
      \frac{\partial L_v}{\partial I_{u}^{N}} +
      \frac{\partial L_v}{\partial {X_{\mu}          }} \frac{d {X_{\mu}          }}{d I_{u}^{N}} +
      \frac{\partial L_v}{\partial {X_{\mu}^{\dagger}}} \frac{d {X_{\mu}^{\dagger}}}{d I_{u}^{N}}
      \right]\Bigg|_{\vec{I}^{N}=0}
      \nonumber
      \\
      & =
      \frac{\gamma^{N}}{S}
      \bigg[
      \Tr\big\{
        \mat{h}^{\text{4c}[I_{u}^{N}]}
        \mat{D}^{\text{4c}[0]}_{v}
      \big\}
      -
      \varepsilon_{v,i}^{4c[0]}
      \Tr\big\{
        \mat{S}^{\text{4c}[I_{u}^{N}]}
        \mat{d}^{\text{4c}[0]}_{v,i}
      \big\}
      \nonumber
      \\
      & +
      \Tr\big\{
        \mat{G}^{\text{4c}[I_{u}^{N}]}
        \mat{D}^{\text{4c}[0]}_{v}
      \big\}
      +
      \Tr\big\{
        \mat{V}^{\text{4c}[I_{u}^{N}]}_{\text{xc}}
        \mat{D}^{\text{4c}[0]}_{v}
      \big\}
      \bigg].
\end{align}
Although the above expression is theoretically valid, its practical application is hindered by the poor convergence of integrals involving one- or two-electron potentials, when evaluated using a Gaussian-type basis set as the basis size increases.~\cite{Ishida2024} To overcome this issue, the Lagrange functional is formulated in operator form (involving MOs rather than MO coefficients), and by applying the Hellmann–Feynman theorem~\cite{Hellmann1937, Feynman1939}, one obtains a formulation of the A-tensor in the RKB basis\cite{Malkin2011}
\begin{align}\label{eq:4c-hfcc}
      A_{uv}^{N}
      =
      \frac{\gamma^{N}}{S}
      \Tr\big\{
        \mat{h}^{\text{4c}[I_{u}^{N}]}
        \mat{D}^{\text{4c}[0]}_{v}
      \big\}.
\end{align}
Similarly, one can obtain the gauge-origin dependent expression for calculating the g-tensor\cite{Repisky2010}
\begin{align}\label{eq:4c-gt-rkb}
      g_{uv}
      =
      \frac{2c}{S}
      \Tr\big\{
        \mat{h}^{\text{4c}[B_{u}]}
        \mat{D}^{\text{4c}[0]}_{v}
      \big\}.
\end{align}
Here, the matrices $\mat{h}^{\text{4c}[I_{u}^{N}]}$ and $\mat{h}^{\text{4c}[B_{u}]}$ contain only an explicit dependence on the nuclear spin and the uniform magnetic field, respectively; that is, the perturbation operators are expressed solely in the RKB basis. Thus, a similar dilemma---whether to use RMB or RKB basis in the formulation of the 4c working equations---also arises for the electronic g-tensor. In this case, the formulation using the RMB basis (without the London phase factor) yields results that are practically identical to those obtained with the RKB basis.\cite{Misenkova2022} In addition, satisfactory convergence with basis set size has been observed for both EPR tensors.\cite{Malkin2011, Misenkova2022} Therefore, the use of the RKB basis for evaluating g-tensor and A-tensor is justified. However, in g-tensor calculations, employing London atomic orbitals in combination with the RMB basis further improves the convergence with the basis set size, particularly in cases involving highly delocalized spin densities.\cite {Misenkova2022}

The expressions in Eqs.~\eqref{eq:4c-hfcc} and~\eqref{eq:4c-gt-rkb} involve the trace of the unperturbed zero-order density matrix with only one-electron integrals. As the 4c density matrix can be expressed in terms of the 2c density matrix and the X2C decoupling matrix $\mat{U}$, one can derive the final amfX2C expressions for the hyperfine coupling tensor and the gauge-origin dependent g-tensor, both in the RKB basis
\begin{align}
      A^N_{uv} 
      & = 
      \frac{\gamma^{N}}{S} 
      \Tr\big\{
        \mat{\tilde{h}}^{\text{2c}[I_{u}^{N}]} 
        \mat{\tilde{D}}^{\text{2c}[0]}_{v}
      \big\},
      \qquad
      \mat{\tilde{h}}^{\text{2c}[I_{u}^{N}]}
      =
      \Big[
      \mat{U}^{\dagger}
      \mat{h}^{\text{4c}[I_{u}^{N}]}
      \mat{U}
      \Big]^{++},
      \\
      g_{uv} 
      & = 
      \frac{2c}{S} 
      \Tr\big\{
        \mat{\tilde{h}}^{\text{2c}[B_{u}]} 
        \mat{\tilde{D}}^{\text{2c}[0]}_{v}
      \big\},
      \qquad
      \mat{\tilde{h}}^{\text{2c}[B_{u}]}
      =
      \Big[
      \mat{U}^{\dagger}
      \mat{h}^{\text{4c}[B_{u}]}
      \mat{U}
      \Big]^{++}.
\end{align}
Due to the simplified formulation, it is sufficient to apply the picture-change transformation only to the one-electron integrals.

\begin{figure}[htb!]
    \centering
    \includegraphics[width=\linewidth]{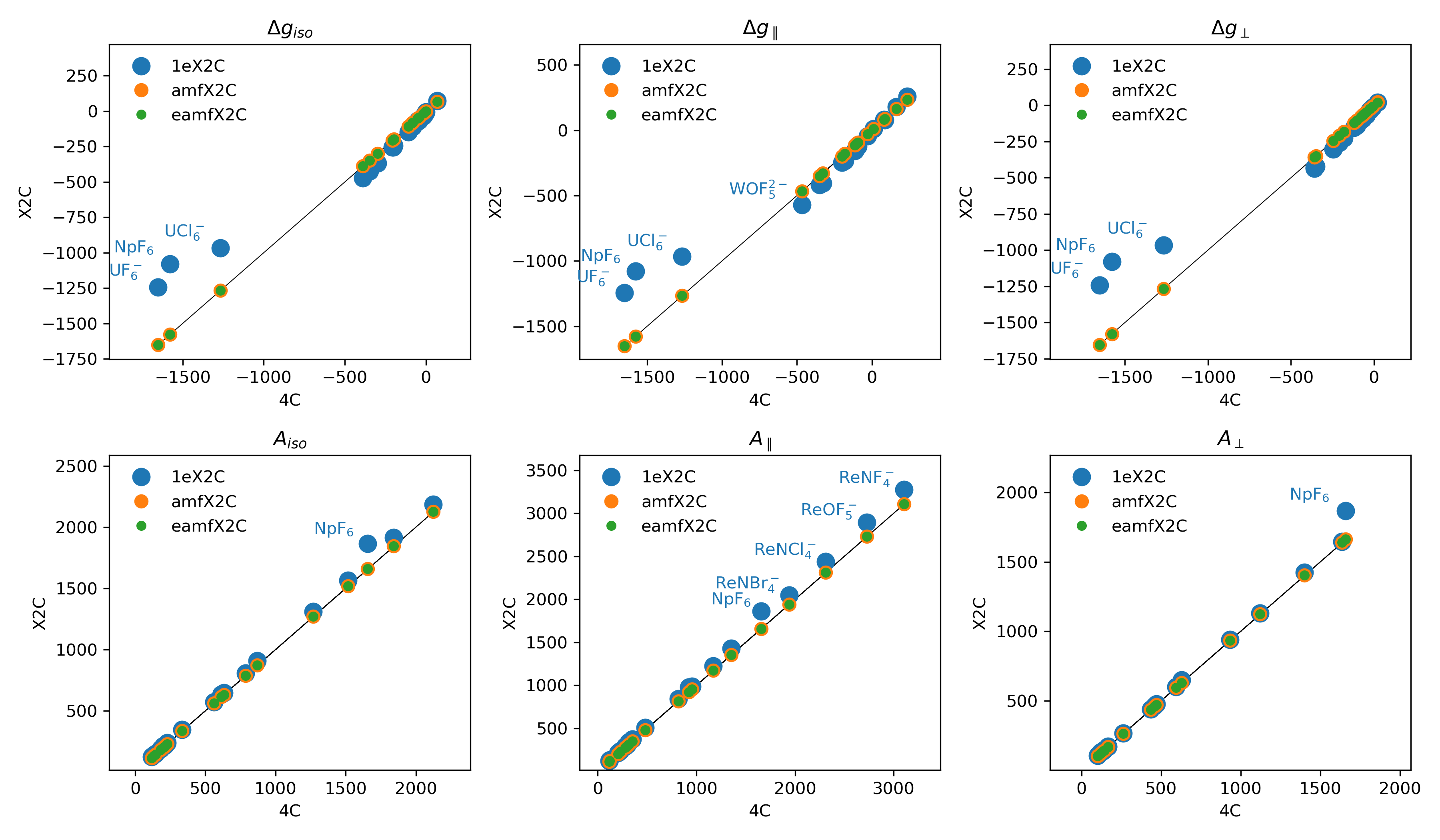}
    \caption{Comparison of isotropic (iso), parallel ($_\parallel$), and perpendicular ($_\perp$) principal components of the EPR $\Delta g$-tensor (in ppt) and metal hyperfine coupling $A$-tensor (in MHz) obtained from RKB/PBE0-40HF/dyall-vtz/IGLO-III calculations for three variants of X2C Hamiltonian models in comparison with reference four-component (4C) Dirac–Coulomb results for a standard benchmark dataset of small complexes. The molecules with the largest deviations from reference 4C data are marked. The complete dataset is reported in Ref.~\citenum{Repisky2025}.}
    \label{fig:epr}
\end{figure}

\begin{table}[]
 \centering
 \begin{tabular}{l*{7}r}
 \hline \hline\noalign{\smallskip}
{Molecule} & {Hamiltonian} & {$\Delta g_{\text{iso}}$} & {$\Delta g_{\parallel}$} & {$\Delta g_{\perp}$} & {$A_{\text{iso}}$} & {$A_{\parallel}$} & {$A_{\perp}$} \\ \hline
WOBr$_5^{2-}$&1eX2C& -2\highlight{53.4}& -1\highlight{53.9}& -\highlight{303.1}& 2\highlight{11.5}& 3\highlight{40.7}& 14\highlight{6.9}\\
&amfX2C&             -201.\highlight{2}& -1\highlight{10.9}& -246.3            & 20\highlight{1.0}& 315.\highlight{8}& 143.\highlight{6}\\
&eamfX2C&            -201.3            & -111.\highlight{4}& -246.3            & 20\highlight{1.0}& 315.\highlight{8}& 143.\highlight{6}\\
&4C&                 -201.3            & -111.5            & -246.3            & 200.7            & 315.4            & 143.3\\
\\
ReOBr$_4$&1eX2C& -\highlight{64.6}& 2\highlight{58.1}& -\highlight{226.0}& \highlight{908.7}& 1\highlight{430.1}& 6\highlight{48.0}\\
&amfX2C        & -42.\highlight{3}& 235.\highlight{5}& -181.2            & 87\highlight{2.1}& 135\highlight{5.1}& 6\highlight{30.6}\\
&eamfX2C       & -42.4            & 235.1            & -181.2            & 87\highlight{2.2}& 135\highlight{5.4}& 6\highlight{30.5}\\
&4C            & -42.4            & 235.1            & -181.2            & 870.5            & 1353.6            & 628.9\\
\\
ReNF$_4^-$&1eX2C& -\highlight{242.2}& -\highlight{417.7}& -1\highlight{54.4}& 21\highlight{87.8}& 3\highlight{273.7}& 16\highlight{44.8}\\
&amfX2C         & -19\highlight{7.0}& -349.\highlight{2}& -120.9            & 212\highlight{9.0}& 310\highlight{8.5}& 163\highlight{9.2}\\
&eamfX2C        & -196.9            & -349.\highlight{0}& -120.9            & 212\highlight{9.0}& 310\highlight{8.3}& 163\highlight{9.3}\\
&4C             & -196.9            & -349.1            & -120.9            & 2125.1            & 3104.2            & 1635.5\\
\\
OsOF$_5$&1eX2C& -\highlight{368.5}& -\highlight{232.6}& -\highlight{436.4}& 6\highlight{33.9}& 9\highlight{77.0}& 4\highlight{62.3}\\
&amfX2C       & -298.\highlight{7}& -179.6            & -358.\highlight{3}& 6\highlight{13.6}& 92\highlight{4.3}& 45\highlight{8.2}\\
&eamfX2C      & -298.5            & -179.\highlight{5}& -358.0            & 61\highlight{3.5}& 92\highlight{4.3}& 45\highlight{8.2}\\
&4C           & -298.5            & -179.6            & -358.0            & 612.4            & 923.1            & 457.1\\
\\
NpF$_6$&1eX2C& -1\highlight{079.3}& -1\highlight{079.4}& -1\highlight{079.2}& 1\highlight{866.1}& 1\highlight{862.5}& 1\highlight{867.9}\\
&amfX2C      & -1577.\highlight{8}& -1577.\highlight{7}& -157\highlight{8.0}& 16\highlight{60.2}& 1655.\highlight{9}& 16\highlight{62.3}\\
&eamfX2C     & -157\highlight{8.0}& -1577.\highlight{9}& -157\highlight{8.1}& 16\highlight{60.1}& 1655.\highlight{9}& 16\highlight{62.2}\\
&4C          & -1577.5            & -1577.4            & -1577.6            & 1657.5            & 1655.0            & 1658.8\\
\\
 \hline \hline
\end{tabular}
\caption{Computed isotropic (iso), parallel ($_\parallel$), and perpendicular ($_\perp$) principal components of the EPR $\Delta g$-tensor (in ppt) and metal hyperfine coupling $A$-tensor (in MHz) obtained from RKB/PBE0-40HF/dyall-vtz/IGLO-III calculations for three variants of X2C Hamiltonian models in comparison with reference four-component (4C) Dirac–Coulomb results for a selected set of complexes from a standard benchmark dataset. The differences in comparison to reference 4C data are marked in red. The complete dataset is reported in Ref.~\citenum{Repisky2025}.
}
\label{tab:epr}
\end{table}

The assessment of the X2C models for EPR properties was performed on a set of 5d$^1$  and an exemplary 5f$^1$  transition metal complexes, \textit{vis-a-vis} WOBr$_5^{2-}$, ReOBr$_4$, ReNF$_4^-$, OsOF$_5$ and NpF$_6$. 
DFT and CASPT2 optimized geometries were used for the 5d and 5f molecules from Ref.~\citenum{Gohr2015} and ~\citenum{Notter2009}, respectively.
All calculations reported in Table~\ref{tab:epr} were performed employing a finite value of speed of light $c= 137.035999074$ a.u., a spherical Gaussian nuclear charge distribution, and a point model for nuclear magnetic moments~\cite{Malkin2011}. 
A previously established computational protocol~\cite{Gohr2015} for relativistic EPR calculations comprising of a customized PBE0-40HF~\cite{Adamo1999}, all-electron uncontracted dyall-vtz basis~\cite{Dyall2004-5d,Dyall-TCA-117-483-2007,Dyall2010-5d} for heavy elements (Br, W, Re, Os, Np) and IGLO-III basis~\cite{Kutzelnigg1990} for light elements (N, O, F) was used. For the reference four-component (4c) calculations, the small-component basis was generated automatically by \ReSpect ~applying the RKB condition to the large-component dyall-vtz/IGLO-III basis. The exchange-correlation contributions were evaluated numerically on \ReSpect’s default integration grid, employing the modified Scalmani–Frish~\cite{Scalmani2012} noncollinear XC parametrization.\cite{Komorovsky2019, Repisky2020}

All EPR parameters were calculated by means of the first-order perturbation theory, facilitating Malkin’s 3SCF approach with three Kramers-unrestricted Kohn-Sham determinants computed for three distinct orthogonal noncollinear spin-magnetizations.~\cite{Malkin2005, Komorovsky2023} Each SCF was converged below $10^{-7}$ in both the DIIS error vector and the change in electronic energy. The hyperfine coupling principal values $A_{ii}, i=1-3$ were obtained as the square roots of the eigenvalues of the matrix $AA^T$, while the principal axis system (PAS) is given by the eigenvectors. The signs of the hyperfine coupling components are reported as calculated and adjusted only, where needed, to either match the LR results or by ensuring that the sign of the determinant of the hyperfine coupling matrix before and after rotation to the PAS remains the same. The isotropic hyperfine coupling is given by $A_{iso}= (1/3) \sum_i A_{ii}$, and we report the hyperfine coupling values parallel ($A_\parallel$) and perpendicular ($A_\perp$) to the principal symmetry axis.

Fig.~\ref{fig:epr} compares the EPR $\Delta g$-tensor and hyperfine $A$-tensor obtained using three X2C variants against reference 4C results. The calculations were carried out on a standard benchmark dataset of small molecules containing 4d, 5d and 5f elements. It is evident that the 1eX2C model exhibits large errors, stemming from the neglect of 2ePC and xcPC corrections. The results for the complexes with the highest deviations are reported in Table.~\ref{tab:epr}, which shows that the $\Delta g$-tensor is more sensitive than the $A$-tensor calculated using the 1eX2C Hamiltonian. For instance, the 1eX2C errors for NpF$_6$ is $\sim 31 \%$ and $\sim 12\%$ for the EPR $\Delta g$-tensor and $A$-tensor, respectively.   
In contrast, both amfX2C and eamfX2C yield dramatic improvements, consistently reducing errors to below $\sim 1\%$ for both $\Delta g$ amf $A$-tensors in comparison to reference 4C. This can be attributed solely to the inclusion of the 2ePC and xcPC corrections based on the amf-based approximations.   
Generally, amfX2C and eamfX2C outperforms 1eX2C 
by at least two orders of magnitude for the $\Delta g$-tensor and by an order of magnitude for the $A$-tensor. A more extensive set of systems is reported in Ref.~\citenum{Repisky2025}.

\vspace{0.4cm}

%%%%%%%%%%%%%%%%%%%%%%%%%%%%%%%%%%%%%%%%%%%%%%%%%%%%%%%%%%%%%%%%%%%%
\subsection{\label{sec:applications-RT}C: X2C Real-Time TDDFT}

Real-time time-dependent density functional theory (RT-TDDFT) offers a direct and non-perturbative approach to studying electron dynamics and spectroscopic processes in molecular systems subjected to external electromagnetic fields. The non-perturbative nature of RT-TDDFT allows for the incorporation of fields with arbitrary intensity, duration, shape, or energy. Compared to response theory, RT-TDDFT accounts for both linear and non-linear effects without the need to implement response kernels or suffer from divergences at resonant frequencies~\cite{Yabana1996,Repisky2015,Li2020}.

In our previous \ReSpect~article~\cite{Repisky2020}, ground-state spectral property simulations in response to a \emph{single} probe pulse were discussed~\cite{Repisky2015,Kadek2015,Konecny2016,Konecny2018}. These calculations were performed at various levels of theory: non-relativistic (1c), two-component (2c) using the 1eX2C Hamiltonian model, and four-component (4c) with the Dirac–Coulomb Hamiltonian. Rather than repeating these results, we focus here on recent developments and highlight extensions of the RT-TDDFT module in \ReSpect{} to non-stationary state dynamics to enable the simulation of \emph{time-resolved pump–probe spectroscopies}~\cite{Moitra2023,Moitra2025}. In these techniques, a pump pulse drives the system out of equilibrium and creates an electronic wavepacket, while a probe pulse captures the wavepacket's response. In contrast to conventional stationary-state dynamics, the features of the pump pulse and the time delay between the pump and probe pulses also strongly influence the spectroscopic signatures.

The simulation of pump–probe spectroscopic processes in \ReSpect{} is based on solving the Liouville--von Neumann (LvN) equation of motion in time domain using Kohn--Sham DFT and an orthonormal basis~\cite{Moitra2023,Moitra2025},
\begin{align}
    i\frac{\partial}{\partial t}\mat{D}(t) &= [\mat{F}(t), \mat{D}(t)].
\end{align}
Here, the one-electron density matrix $\mat{D}(t)$ describes the state of the molecular system at time $t$, while the Fock matrix $\mat{F}(t)$ characterizes both the system itself and its interaction with external time-dependent driving fields. To simulate pump–probe experiments, 
we introduce two classical electric fields within the long-wavelength approximation: 
$\vec{\mathcal{E}}(t)$ for the pump and $\vec{\mathcal{F}}(t)$ for the probe. These fields are coupled to the molecular system via the electric dipole moment matrix, $\mat{P} = (\mat{P}_{x}, \mat{P}{y}, \mat{P}_{z})$. The resulting Fock matrix can be expressed as~\cite{Moitra2023,Moitra2025}
\begin{align}
   \label{eq:Fock}
   \mat{F}(t)
   := 
   \mat{F}^0[\mat{D}(t)]
   -
   \sum_{k\in x,y,z}
   \mat{P}_{k} \mathcal{E}_{k}(t)
   -
   \sum_{k\in x,y,z}
   \mat{P}_{k} \mathcal{F}_{k}(t)
   ,
\end{align}
where $\mat{F}^0[\mat{D}(t)]$ represents the field-free Fock matrix. Within a finite time window, solving the LvN equation reduces to evaluating the time-dependent Fock matrix at discrete time steps and propagating the density matrix in time. In \ReSpect{}, numerically stable propagation is achieved using a second-order Magnus propagator~\cite{Magnus1954} combined with self-consistent microiterations~\cite{Repisky2015}.

Currently, \ReSpect~allows the choice of both linearly-polarized (LP) and circularly polarized (CP) pump pulse by tailoring the parameters in the following expression: 
\begin{align}
\vec{\mathcal{E}}(t) &= 
%\mathcal{E}_{0} \sum_{i=1,2}\vec{e}_{i}f(t-t_0,\omega_{0}, \phi_i)
%&=
\begin{cases}
    \mathcal{E}_0  
    f(t; t_0,T)\bm{e}_1
    %\cos^2 \left(\pi \frac{t-t_0}{T} \right)	
    \sin[\omega_0 (t-t_0) ] 
    & \text{LP}  
    \\
    \mathcal{E}_{0} f(t; t_0,T) \big(\bm{e}_1\cos[\omega_{0}(t-t_{0})] 
    +
    \bm{e}_2\cos[\omega_{0}(t-t_{0})-\pi/2] \big) & \text{CPR} \\
    \mathcal{E}_{0}f(t; t_0,T) \big(\bm{e}_1\cos[\omega_{0}(t-t_{0})] 
    +
    \bm{e}_2\cos[\omega_{0}(t-t_{0})+\pi/2]\big) & \text{CPL}
\end{cases}
\label{eq:pump}
\end{align}
In the above expression, $\mathcal{E}_{0}$, $f$ and $\vec{e}_1 \perp \vec{e}_2$ represents the pulse amplitude, a real time-dependent envelope function, and polarization unit-vectors, respectively. In addition, $t_0$, $\omega_0$, and $T$ denote the pulse center, carrier frequency, and duration, respectively. For LP and CP pulses, typical choices are the $\cos^2$ and Gaussian envelope functions, respectively. As the probe pulse, we apply a weak delta-type, 
\begin{align}
    \vec{\mathcal{F}}(t)&= \mathcal{F}_0\vec{e}_1\delta(t-(T+\tau)),
    \label{eq:probe}
\end{align}
with amplitude $\mathcal{F}_0$($<\mathcal{E}_0$), polarization direction $\vec{e}_1$, and applied at time $T+\tau$, where $\tau$ is the time delay between pump and probe pulses,

\begin{figure}
    \centering
    \includegraphics[width=0.9\textwidth]{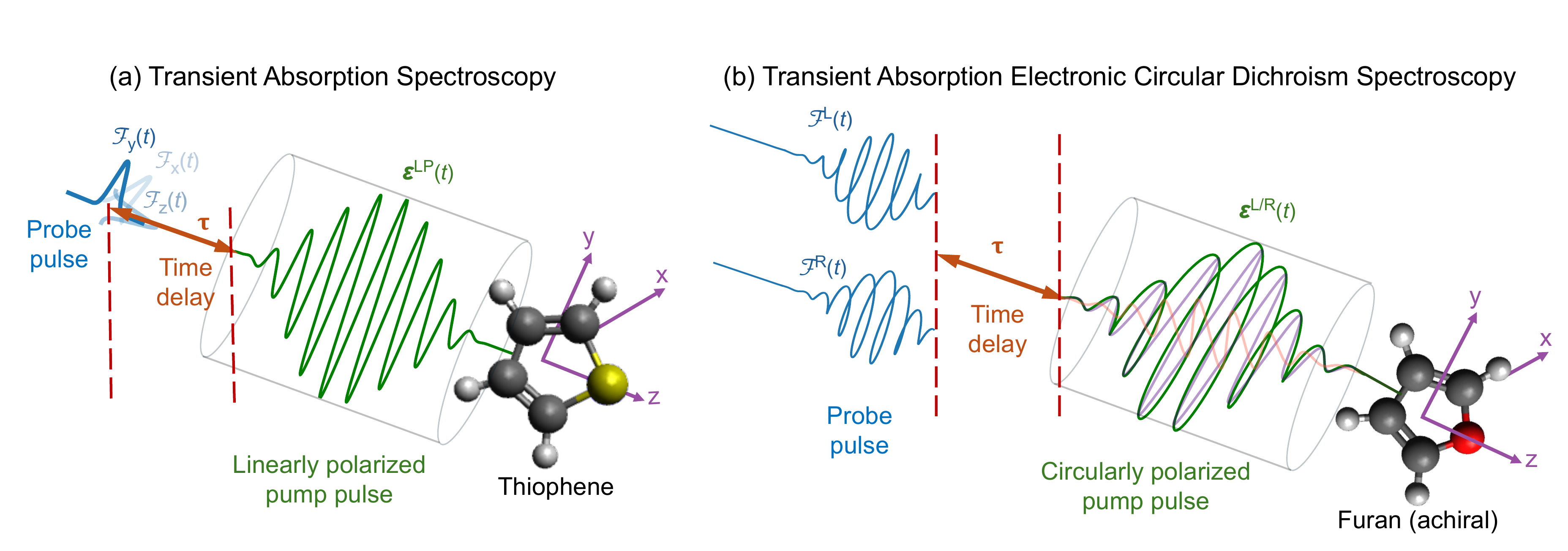}
    \caption{Pump-Probe setup for (a) transient absorption spectroscopy and (b) transient absorption electronic circular dichroism. The mathematical expressions for the pulses are given in Eqs.~\eqref{eq:pump} and \eqref{eq:probe}. }
    \label{fig:RT-setup}
\end{figure}

\subsubsection{Transient Absorption Spectroscopy}
\begin{figure}
    \centering
    \includegraphics[width=0.9\textwidth]{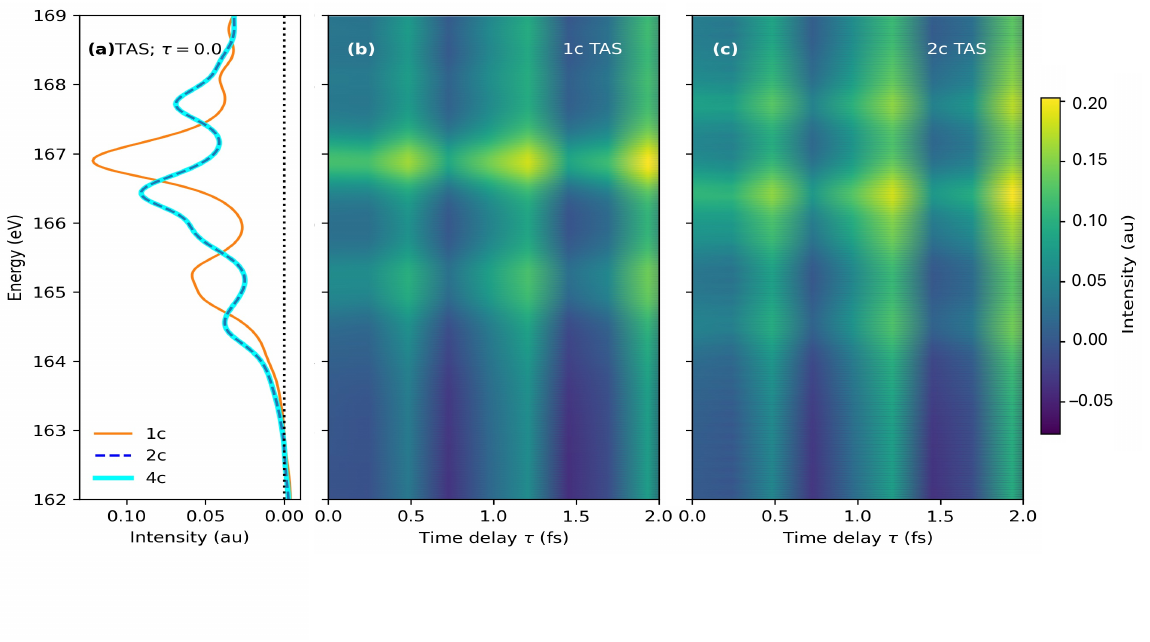}
    \caption{Transient absorption spectra of thiophene:  (a) At time-delay $\tau = 0$ with 1c KS (orange), 2c amfX2C (blue), and 4c DC (cyan) Hamiltonians. Variation in TAS spectra with $\tau$ obtained with the (b) 1c KS and (c) 2c amfX2C Hamiltonian. The figure is reproduced from Ref.~\citenum{Moitra2023}.
    }
    \label{fig:TAS}
\end{figure}

As a first example, we discuss pump-probe transient absorption spectroscopy at the sulfur $L$-edge of thiophene. Although sulfur is a relatively light element, it exhibits pronounced relativistic effects in the core region, as will be demonstrated here. 
Here, we use a LP pump pulse, tuned to the first excited state, to create a non-stationary wavepacket. After a time-delay, a LP delta-type probe pulse captures the absorption spectral imprint of the non-stationary state, as shown in Fig.~\ref{fig:RT-setup}a. 
During the RT-propagation, we record the induced electric dipole moment of the system, which is linked to the absorption spectral function. 

The corresponding X-ray transient absorption spectrum is shown Fig~\ref{fig:TAS}. In the left-most panel, we compare the results at the non-relativistic (1c), amfX2C (2c) and Dirac-Coulomb (4c) levels when the pump-probe time-delay $\tau=0$. 
The main observations are as follows:
(i) As anticipated, the non-relativistic calculation shows only two spectral peaks corresponding to excitation from $2s$ and $2p$ orbitals of sulphur. In contrast, the relativistic simulations yield a richer spectrum, with a scalar shift of the $2s$ ($L_1$) feature and SOC driven splitting of the $2p$ feature into $L_2$ and $L_3$ peaks. 
(ii) The spectral functions obtained with amfX2C and the reference Dirac-Coulomb Hamiltonian exhibit excellent agreement, while the former is an order of magnitude faster than the reference method. 
(iii) The spectral functions have negative intensity features, which are a hallmark of the non-stationary state being probed. These features have been better understood from the point-of-view of non-equilibrium response theory.~\cite{Moitra2023}

Fig.~\ref{fig:TAS}b,c shows the effect of varying the time-delay between pump and probe pulses. In essence, this captures the dynamic evolution of the electronic wavepacket at the time of application of the probe. 
Comparing the non-relativistic and amfX2C spectral functions reveals the same characteristics as discussed above for $\tau=0$. Note, that we are looking at only the first few femtoseconds after end of pump pulse, where nuclear degrees of freedom are frozen and the dynamics is entirely governed by electronic motion. 

In short, the take-away from these results is that the amfX2C method can be used with confidence to obtain 4C quality spectral functions, even for larger systems due to its reduced computational cost. For further details about the computational setup and spectral analysis we refer the readers to Ref.~\citenum{Moitra2023}.

\subsubsection{Transient Absorption Electronic Circular Dichroism}

\begin{figure}
    \centering
    \includegraphics[width=0.7\textwidth]{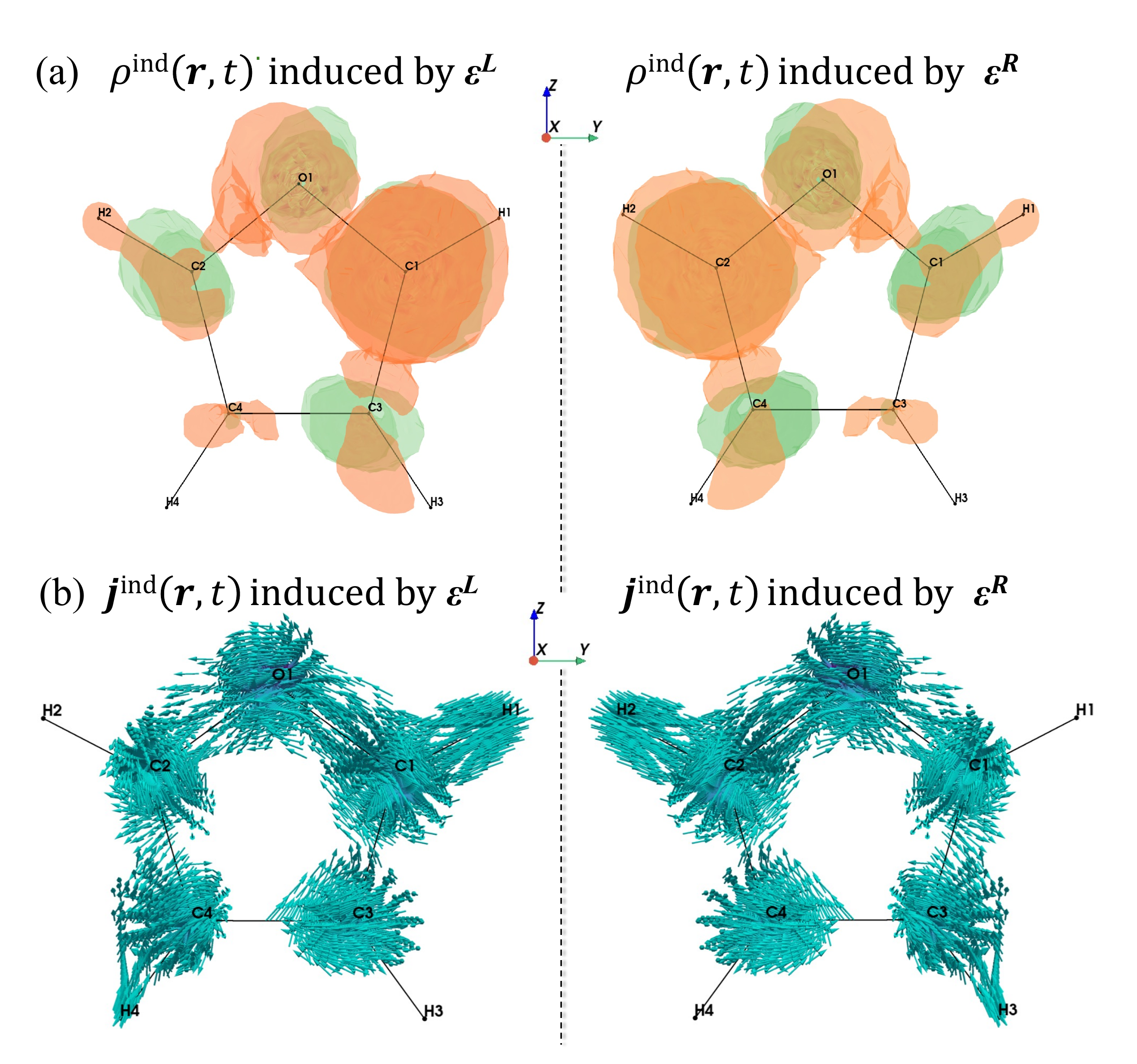}
    \caption{Circularly polarized left (L) and right (R) pump pulse induced (a) charge and (b) current densities captured at the end of pump pulse. }
    \label{fig:RT-densities}
\end{figure}

\begin{figure}
    \centering
    \includegraphics[width=0.9\textwidth]{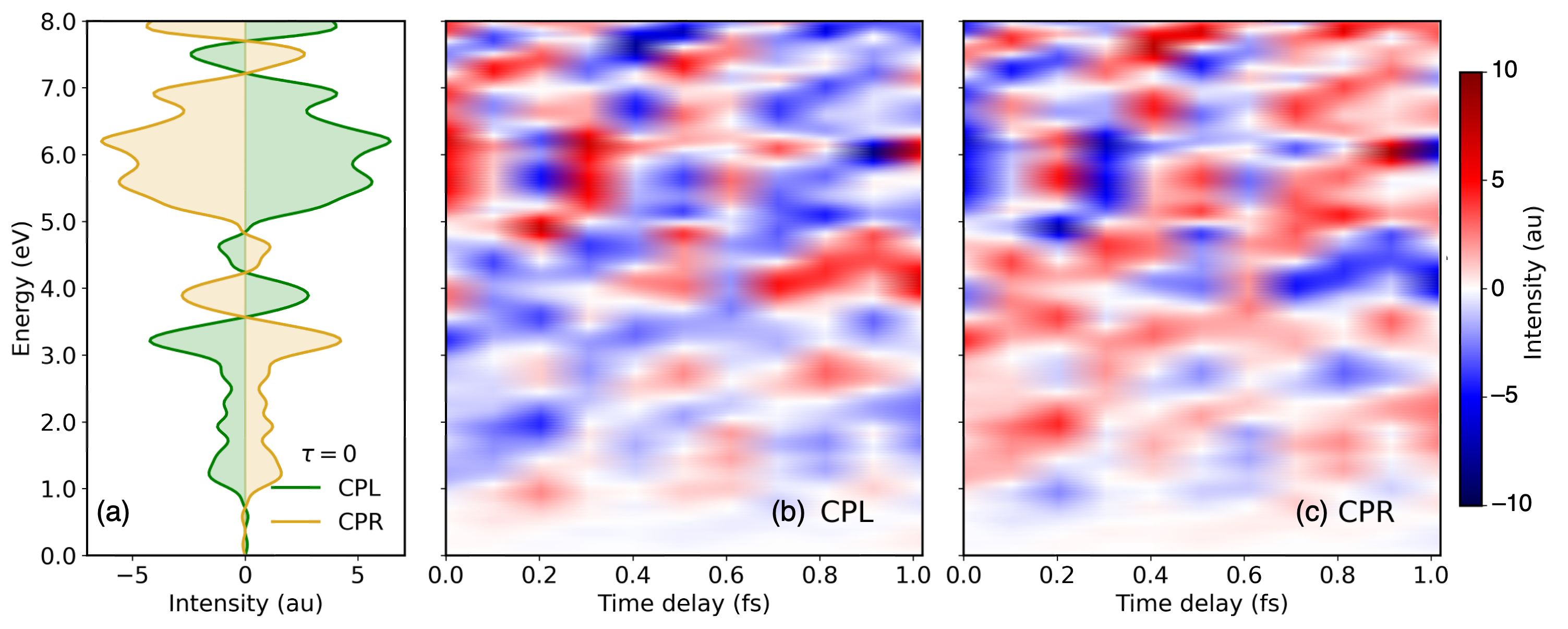}
    \caption{Transient absorption electronic circular dichroism spectra of aligned furan:  (a) At time-delay $\tau = 0$ with CPL (green) and CPR (yellow) pump pulses. Variation in TAS spectra with $\tau$ obtained with the (b) CPL and (c) CPR pump pulses. The figure is reproduced from Ref.~\citenum{Moitra2025}.}
    \label{fig:TRECD}
\end{figure}

Next, we show exemplary results for simulating time-resolved electronic circular dichroism signals by applying a chiral CP pump to an oriented achiral furan molecule. The computational pump-probe setup is shown in Fig.~\ref{fig:RT-setup}b.
This is a proof-of principle example involving a light element containing molecule, in the UV-vis energy region. Hence, only non-relativisitc results are discussed. 

In order to get a microscopic view of the electronic motion, we have added a new functionality in \ReSpect ~to visualise the induced charge and current density during the RT-propagation. For instance, Fig.~\ref{fig:RT-densities} shows the charge and current density induced after the end of CP pump pulse interacting with aligned furan. It clearly shows chiral mirror-image symmetric distributions of electron charge and current densities resulting from the light-matter interaction, depending on the polarization of the light. More importantly, this feature enables a deeper understanding of the mechanism underlying the observed spectroscopic signatures.

Further, the induced current density gives rise to a corresponding induced magnetic dipole moment in the system, which is recorded at each time-step of RT simulation.  
A standard approach to spectroscopically detect the induced magnetic dipole moment in structurally chiral systems is through electronic circular dichroism (ECD) absorption spectroscopy. 
Theoretically, the ECD spectral function is calculated in the weak-probe regime from the imaginary part of the Rosenfeld tensor. 
In the RT-TDDFT framework, this tensor can be obtained by the Fourier transformation of the time-dependent induced magnetic dipole moment recorded during time propagation.  We have extended this technique to a typical pump-probe setup for obtaining time-resolved (TR-)ECD spectral signature.
The technical details was discussed extensively in Ref.~\citenum{Moitra2025,Repisky2020,Konecny2018}. Note that this is also referred to as transient absorption electronic circular dichroism spectra, as experimentally it is a measure of differential absorption of left and right CP light. 

In this case, the evolution of the chiral electronic wavepacket is monitored by varying the time-delay between pump and probe pulses, and recording the spectral function at various time-delays, as shown in Fig.~\ref{fig:TRECD}. The spectra are rich in information, the significant observations are as follows:  (i) The mirror-image symmetry between CPL and CPR induced spectral signal at all time-delays have a mirror-image relationship. This indicates that the enantiomeric character of the induced wavepackets are preserved during time evolution. (ii) The sign reversals in the spectral signal does not occur at a single characteristic time throughout the energy span. These findings and the methodology is described in details in Ref.~\citenum{Moitra2025}.

% \subsubsection{Summary}
% The RT-TDDFT electron dynamics module in \ReSpect ~allows the calculation of the following molecular stationary and non-stationary properties:~\cite{Repisky2015,Kadek2015,Konecny2018,Moitra2023,Kadek2024,Moitra2025}
% \begin{itemize}
%     \item ground state and time-resolved electronic absorption spectra
%     \item ground state and time-resolved electronic circular dichroism spectra
%     \item ground state and time-resolved optical rotatory dispersion spectra
%     \item induced charge and current density visualization
% \end{itemize}
% The current list of features includes,
% \begin{itemize}
%     \item 1c(KS), 2c (amfX2C and 1eX2C), and 4c (DC) Hamiltonians \fixme{pump-probe 1eX2C not tested}
%     \item possibility of addressing arbitrary spectral regions (uv-vis and x-ray) for all stationary state spectra and TAS. \fixme{TCD is not tested for core}
%     \item possibility of applying perturbation/response operators only to a selected molecular orbital window
%     \item flexibility to choose pump pulse features for pump-probe setups
%     \item evaluation of spectra by means of Fourier transform
%     \item visualization of induced charge and current density 
%     \item restart from previous calculations
%     \item resolution-of-the-identity integral acceleration for the Coulomb term (RI-J) and
%     \item hybrid MPI/OpenMP parallelization.
% \end{itemize}

%%%%%%%%%%%%%%%%%%%%%%%%%%%%%%%%%%%%%%%%%%%%%%%%%%%%%%%%%%%%%%%%%%%%
\subsection{\label{sec:applications-LR}D: X2C Linear-Response TDDFT}

For many applications, it is sufficient and often more efficient to solve the
EOMs of TDDFT by means of perturbation theory, particularly in the linear
response regime.~\cite{Norman2018-book} Such methods are a mainstay of computational quantum
chemical packages and are used to calculate various molecular properties
including excitation energies, transition dipole moments and densities,
absorption and circular dichroism spectra in various domains, $C_6$ dispersion coefficients, and
radiative lifetimes.~\cite{Dreuw2005}

Equations of linear response TDDFT under the X2C transformation can be derived from
the 4c time dependent KS equations by decoupling the equation still in the time domain
followed by the time dependent perturbation theory on the decoupled equation.
The decoupling of the real-time equation was presented in Ref.~\citenum{Konecny2016}
at the level of 1eX2C and further extended into the linear response regime and
more advanced X2C models (i.e. mmfX2C, amfX2C, eamfX2C) in Ref.~\citenum{Konecny2023}.
While the X2C decoupling matrix $\mat{U}(t,\vec{\mathcal{E}})$ in general depends
on time (in the real-time regime and consequently on frequency in the frequency
domain) and external field $\vec{\mathcal{E}}$, it is possible to neglect these
dependencies if the electric dipole approximation and weak field approximations
are considered since it can be shown that they lead to $\partial_t \mat{U}(t,\vec{\mathcal{E}}) \approx 0$.
This regime which we dubbed adiabatic X2C transformation results
in $\mat{U}(t,\vec{\mathcal{E}}) \approx \mat{U}(0,0) \equiv \mat{U}$, meaning
that the picture change transformation in the post SCF calculations is performed
by the same X2C decoupling matrix as in the SCF procedure.
Furthermore, the final working equations of linear response TDDFT under the adiabatic
X2C approximation assume the same form as their 4c counterparts with no additional
terms allowing the use of the same solvers as well as achieving code economy.

The adiabatic X2C decoupled EOM takes the form
\begin{equation}
\label{eq:x2c_eom}
i \partial_t \tilde{C}_{\mu j}^{\mathrm{2c}}(t,\vec{\mathcal{E}})
=
\tilde{F}^{\mathrm{2c}}_{\mu\nu}\left(t,\vec{\mathcal{E}}\right) \tilde{C}_{\nu j}^{\mathrm{2c}}(t,\vec{\mathcal{E}})
,
\end{equation}
where the Fock matrix $\mat{\tilde{F}}^\mathrm{2c}$ is defined under the corresponding
X2C model while also containing an interaction with a harmonic external electric field $\vec{\mathcal{E}}(t)$,
and the transformed MO coefficients are
\begin{equation}
       \tilde{C}^{\mathrm{2c}}_{\mu i}(t,\boldsymbol{\mathcal{E}})
   =
   \Big[ 
        \mathbf{U}^{\dagger} \mathbf{C}^{\mathrm{4c}}(t,\boldsymbol{\mathcal{E}})
   \Big]^{++}_{\mu i}
   .
\end{equation}
In response theory, these coefficients are expanded in the powers of the external field
\begin{align} 
  \label{eq:C2c-epansion}
  \tilde{C}^\mathrm{2c}_{\mu i}(t,\vec{\mathcal{E}})
  & = 
  \sum_{p\in(+)}
  \tilde{C}^\mathrm{2c}_{\mu p} 
  \left[
     \delta_{pi}
     +
     d_{u,pi}^{(1)}(t)
     \mathcal{E}_u 
     +
     O(|\boldsymbol{\mathcal{E}}|^2)
  \right]
  e^{-i\varepsilon_{i}t}
  ,
\end{align}
where $d^{(1)}_{u,pi}(t)$ are the first-order expansion coefficients that are further parametrized as
\begin{equation}
  \label{eq:rspXYansatz}
  \mat{d}^{(1)}(t)
  = 
  \mat{X} e^{- i \omega t + \gamma t} + \mat{Y}^\ast e^{i \omega t + \gamma t}
  ,
\end{equation}
where the undetermined coefficients $\mat{X}$ and $\mat{Y}$ are the final unknowns of linear response
theory and we omitted the Cartesian index $u$ arising from the direction of the external field for clarity.
The working equations used to determine $\mat{X}$ and $\mat{Y}$ are obtained by inserting the expansion
\eqref{eq:C2c-epansion} with the ansatz \eqref{eq:rspXYansatz} into Eq.~\eqref{eq:x2c_eom} and collecting
the first order terms.
The final equation has the algebraic form
\begin{equation}
\label{eq:DR-TDDFT}
\left[
\begin{pmatrix}
\mat{A}^\mathrm{2c}       & \mat{B}^\mathrm{2c}       \\
\mat{B}^{\mathrm{2c}\ast} & \mat{A}^{\mathrm{2c}\ast}
\end{pmatrix}
- (\omega+i\gamma)
\begin{pmatrix}
\mat{1} & \mat{0} \\
\mat{0} & -\mat{1}
\end{pmatrix}
\right]
\begin{pmatrix}
\mat{X} \\
\mat{Y}
\end{pmatrix}
=
\begin{pmatrix}
\tilde{\mat{P}}^\mathrm{2c}       \\
\tilde{\mat{P}}^{\mathrm{2c}\ast}
\end{pmatrix}
,
\end{equation}
where $\omega$ and $\gamma$ are user-defined parameters specifying the external
electric field frequency and a common relaxation (damping) parameter modelling
the finite lifetime of the excited states that leads to finite-width peaks.
The matrices $\mat{A}^\mathrm{2c}$ and $\mat{B}^\mathrm{2c}$ are matrix representations of the response kernel
in the canonical MO basis
\begin{subequations}
\begin{align}
  \label{eq:Aterm}
  A^\text{2c}_{ai,bj} 
  &= 
  \omega_{ai}\delta_{ab}\delta_{ij} 
  +
  \left( g^\text{2c}_{\mu\nu,\kappa\lambda} + k^\mathrm{xc}_{\mu\nu,\kappa\lambda} \right)
  \tilde{C}^{\text{2c}\ast}_{\mu a} \tilde{C}^\text{2c}_{\nu i}
  \tilde{C}^{\text{2c}\ast}_{\kappa j} \tilde{C}^\text{2c}_{\lambda b}
  ,
  \\
  \label{eq:Bterm}
  B^\text{2c}_{ai,bj} 
  &=
  \left( g^\text{2c}_{\mu\nu,\kappa\lambda} + k^\mathrm{xc}_{\mu\nu,\kappa\lambda} \right)
  \tilde{C}^{\text{2c}\ast}_{\mu a} \tilde{C}^\text{2c}_{\nu i}
  \tilde{C}^{\text{2c}\ast}_{\kappa b} \tilde{C}^\text{2c}_{\lambda j}
  ,
  \\
  \label{eq:Kterm}
  \mathbf{k}^\mathrm{xc} &= \mathbf{k}^\mathrm{xc}
       \Big( \boldsymbol{\Omega}^\text{2c}, \mathbf{\tilde{D}}^\text{2c} \Big)
  ,
\end{align}
\end{subequations}
where, $\mathbf{g}^\text{2c}$ are the 2c
untransformed two-electron integrals, and $\mathbf{k}^\mathrm{xc}$ is the
exchange--correlation kernel constructed from 2c untransformed overlap
disribution functions $\boldsymbol{\Omega}^\text{2c}$ and the transformed 2c
density matrix $\mathbf{\tilde{D}}^\text{2c}$.
The DFT exchange--correlation kernel is formulated in a non-collinear
fashion~\cite{Komorovsky2019,Konecny2019,Repisky2020}.
The right-hand side of Eq.~\eqref{eq:DR-TDDFT} contains the picture-change transformed electric dipole moment matrix describing the interaction with the external electric field
\begin{equation}
   \label{eq:X2Cmmf-F}
   \tilde{P}^{\mathrm{2c}}_{u,\mu\nu}
   =
   \Big[
       \mat{U}^{\dagger} \mat{P}^{\mathrm{4c}}_{u} \mat{U}
   \Big]^{++}_{\mu\nu}
   ,\qquad
   P_{u,\mu\nu}^{\mathrm{4c}}
   = 
   -
   \int
   \mat{X}_{\mu}^{\text{4c}\dagger}(\vec{r})(r_{u}-R_{u}) \mat{X}^{\text{4c}}_{\nu}(\vec{r})
   d\vec{r}
   .
\end{equation} 
Eq.~\eqref{eq:DR-TDDFT} is referred to as the damped response (DR) TDDFT equation due to the presence
of the damping parameter $\gamma$ (but can also be found in the literature under the names
Sternheimer equation or complex polarization propagator~\cite{Norman2001}).

Furthermore, we can consider the corresponding homogeneous equation to Eq.~\eqref{eq:DR-TDDFT} that after algebraization takes the form of an eigenvalue (EV) equation (also known as the Casida equation)~\cite{Casida1995, Casida2009, Ullrich2011},
\begin{equation}
\label{eq:EV-TDDFT}
\begin{pmatrix}
\mat{A}^\mathrm{2c}       & \mat{B}^\mathrm{2c}       \\
\mat{B}^{\mathrm{2c}\ast} & \mat{A}^{\mathrm{2c}\ast}
\end{pmatrix}
\begin{pmatrix}
\mat{X}_n \\
\mat{Y}_n 
\end{pmatrix}
=
\omega_n
\begin{pmatrix}
\mat{1} &  \mat{0} \\
\mat{0} & -\mat{1} 
\end{pmatrix}
\begin{pmatrix}
\mat{X}_n \\
\mat{Y}_n
\end{pmatrix}
,
\end{equation}
where the eigenvalue $\omega_n$ is the $n$-th excitation energy of the system.
The equation can be simplified by considering the Tamm--Dancoff approximation (TDA)~\cite{Hirata1999}
also available in \ReSpect{} at the X2C level of theory.

Both Eqs.~\eqref{eq:DR-TDDFT} and \eqref{eq:EV-TDDFT} have the same form in the 4c
and 2c regimes meaning that algorithms developed for their solution at the 4c level
can be directly transferred to the X2C case. Specifically, due to the size of the kernel matrices
$\mat{A}^\mathrm{2c}$ and $\mat{B}^\mathrm{2c}$ that are of dimension
$N_\mathrm{vir}N_\mathrm{occ} \times N_\mathrm{vir}N_\mathrm{occ}$,
the DR and EV-TDDFT equations cannot be solved by direct inversion or elimination methods
for many systems of chemical interest and iterative solutions have to be developed instead.~\cite{Davidson1975, Olsen1988, Olsen1990}
The algorithms implemented in \ReSpect{} are iterative subspace solvers that parametrize the
unknown vectors as linear combinations of trial vectors. The details of these iterative
solvers are discussed in Refs.~\citenum{Konecny2019, Komorovsky2019} and for DR-TDDFT
and EV-TDDFT, respectively.

The most common application of linear response TDDFT is the calculation of electronic absorption
spectra (EAS) defined as the dipole strength function
\begin{equation}
\label{eq:rspFunction}
S(\omega)
=
\frac{4\pi\omega}{3 c} \Im \mathrm{Tr} \left[ \mat{\alpha}(\omega) \right]
,
\end{equation}
where $\mat{\alpha}(\omega)$ is the frequency dependent complex polarizability tensor.
This tensor can be calculated from the results of both DR- and EV-TDDFT.~\cite{Konecny2023}
Specifically, in a DR-TDDFT calculation it is obtained as
\begin{equation}
\label{eq:polarisability-DR}
\alpha_{uv}(\omega)
=
X_{ai,v}(\omega) \tilde{P}^{\mathrm{2c}}_{ia,u} + Y_{ai,v}(\omega) \tilde{P}^{\mathrm{2c}}_{ai,u}
,
\end{equation}
and in an EV-TDDFT calculation via
\begin{equation}
\label{eq:Polarizability-EV}
  \alpha_{uv}(\omega)
  =
  \sum_n
  \left[
  \frac{t_{u,n}^\ast t_{u,v}}{\omega+\omega_n+i\gamma}
  -
  \frac{t_{u,n} t_{u,v}^\ast}{\omega-\omega_n+i\gamma}
  \right]
  ,
\end{equation}
where the transition dipole moments between the ground and excited states are
\begin{equation}
\label{eq:transitionDipoleEV}
t_{u,n}
=
X_{ai,n}(\omega) \tilde{P}^{\mathrm{2c}}_{ia,u} + Y_{ai,n}(\omega) \tilde{P}^{\mathrm{2c}}_{ai,u}
.
\end{equation}
In Eq.~\eqref{eq:polarisability-DR}, a damping parameter $\gamma$ was added in the post-processing step
to turn the line spectra into band spectra.
These procedures are equivalently used to obtain absorption spectra in both UV/Vis and X-ray regions.
Moreover, the transition and response vectors can be analyzed to provide information about the nature
of spectral lines in the language of transitions between ground-state molecular orbitals as well as
to calculate transition densities that can be visualized.

Particularly X-ray absorption (XAS) spectra of heavy metal complexes provide an
illustration of the capabilities and advantages of relativistic linear response
TDDFT approaches.~\cite{Norman2018} Because such complexes often contain tens of
atoms, they can be computationally costly to handle by 4c methods. However,
strong spin--orbit coupling and the resulting splitting of core orbitals
appearing when studying XAS spectra near L and M edges, together with prominent
shifts caused by scalar relativistic effects, mandate an accurate treatment of
relativistic effects. This combined demand for accuracy and efficiency makes
X2C-based approaches particularly attractive in this context. This is especially
true for more advanced X2C Hamiltonians, as the 1eX2C is known to overestimate
SO splitting. In cases involving the large SO splitting typical of heavy-element
core orbitals, this can result in errors on the order of tens of eV. Examples of
such effects are shown in Table~\ref{tab:XAS} (data from
Ref.~\citenum{Konecny2023}) that presents XAS edge positions of various metals
in different compounds. While 1eX2C deviates from the reference 4c data, amfX2C,
eamfX2C (giving identical results as amfX2C), and mmfX2C reproduce these 4c
reference data nearly exactly and thus are also well positioned to reproduce and
explain experimental spectra. Furthermore, to reproduce experimental spectra, we
increased the amount of exact exchange in the hybrid functional based on PBE0 to
a number listed in Table~\ref{tab:XAS}. This is standard practice in DFT
modelling of XAS spectra and is related to the high-density limit of
DFT.~\cite{Levy1985, Levy1989, Levy1991} Importantly, the X2C approaches also
offer 7-9 times speedup when compared to the 4c calculations.

\begin{figure}[h]
\centering
\caption{
XAS spectra of \ce{[W Cl4 (PMePh2)2]} calculated using 4c, 1eX2C, and amfX2C DR-TDDFT and EV-TDDFT.
}
\label{fig:XAS}
  \begin{subfigure}{0.20\textwidth}
    \includegraphics[width=\textwidth]{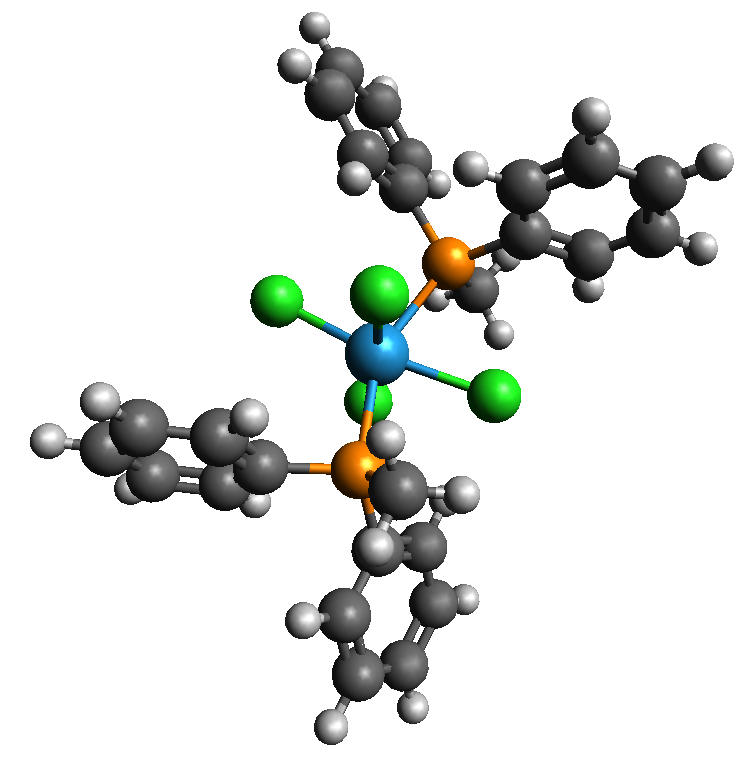}
    \caption{Structure}
    \label{fig:W_molecule}
  \end{subfigure}
  \begin{subfigure}{0.39\textwidth}
    \includegraphics[width=\textwidth]{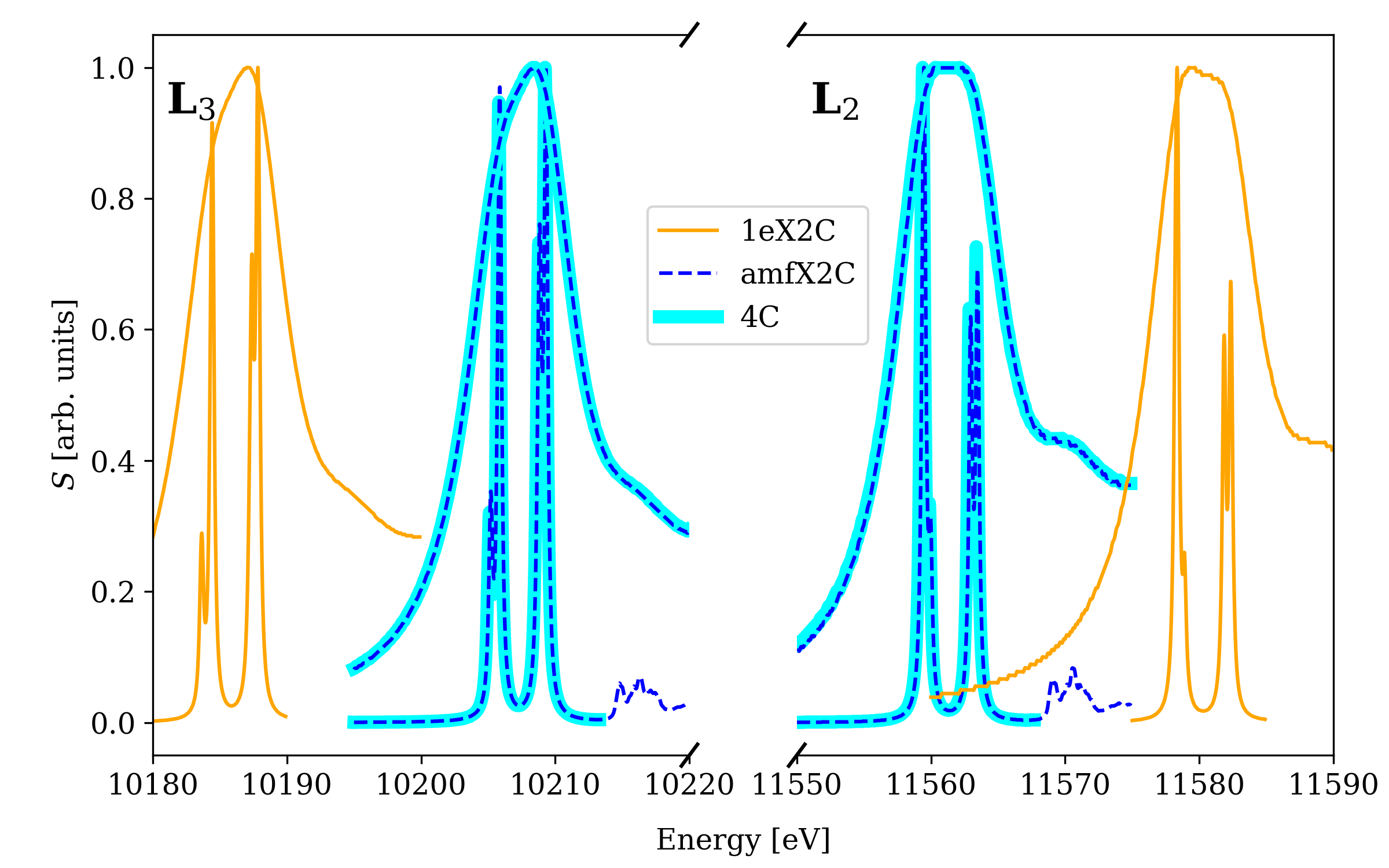}
    \caption{DR-TDDFT}
    \label{fig:W_complex_CPP}
  \end{subfigure}
  \begin{subfigure}{0.39\textwidth}
    \includegraphics[width=\textwidth]{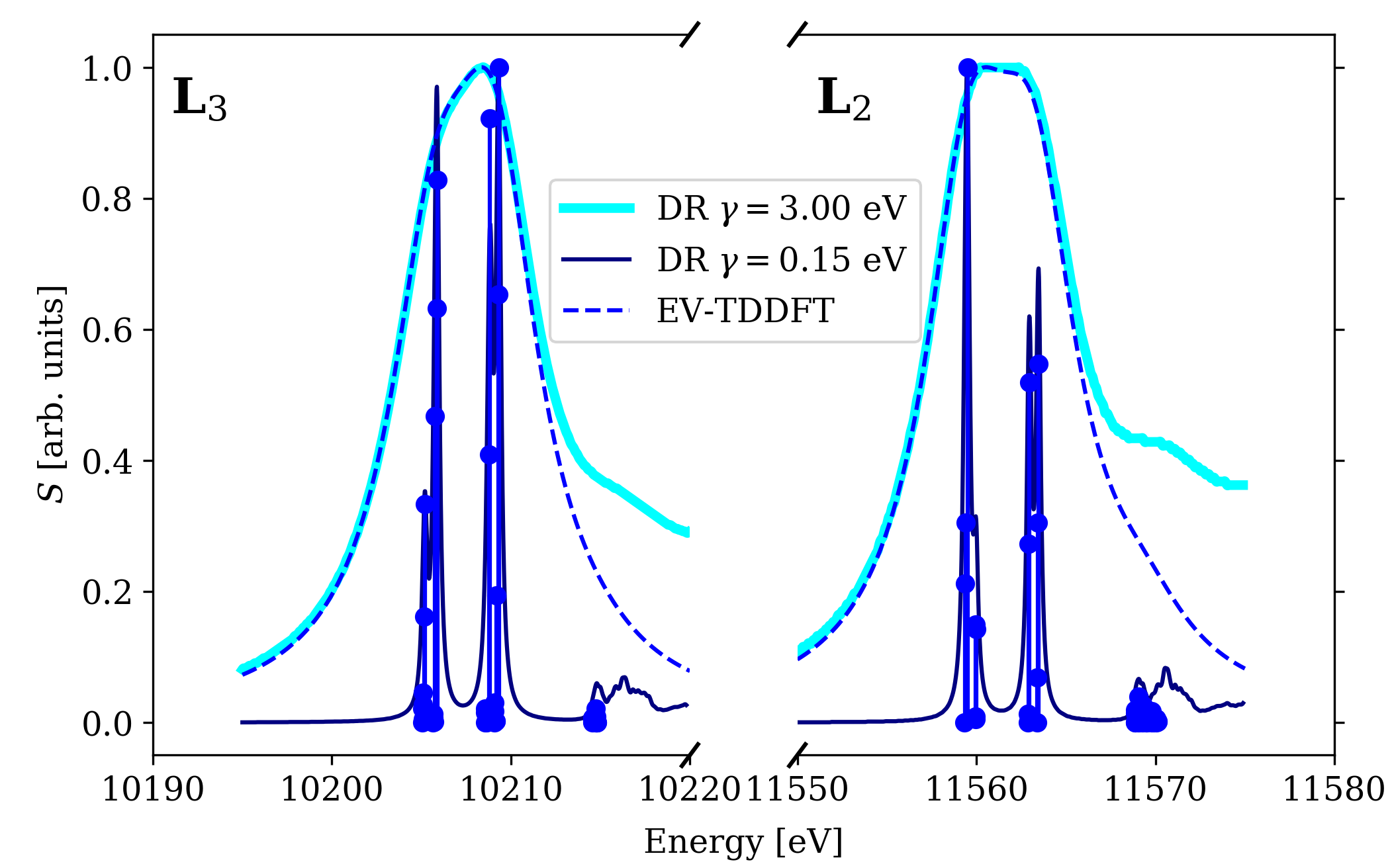}
    \caption{EV-TDDFT}
    \label{fig:W_complex_EV}
  \end{subfigure}
  \label{fig:W_complex_XAS}
\end{figure}

Furthermore, let us examine the \ce{L_{2,3}} edges of \ce{[W Cl4 (PMePh2)2]} where Ph stands for phenyl
(see Fig.~\ref{fig:W_molecule}), a larger molecular system studied by us theoretically in
Refs.~\citenum{Konecny2022} and \citenum{Konecny2023} with experimental data available
in Ref.~\citenum{Jayarathne2014}. We calculated its spectra by means of 4c and 1eX2C and amfX2C
DR-TDDFT shown in Fig.~\ref{fig:W_complex_CPP} and amfX2C EV-TDDFT depicted in Fig.~\ref{fig:W_complex_EV}.
We see that, in this case as well, the 1eX2C approach overestimates the SO splitting by approximately $\unit[40]{eV}$, while the
amfX2C approach reproduces the 4c spectra exactly. The spectra are evaluated with two different values
of the damping parameter, $\gamma = \unit[3.0]{eV}$ reproducing the experimental spectra
and $\gamma = \unit[0.15]{eV}$ to allow a better resolution of the broad bands.
While this approach already allows us to associate the spectral lines with transitions between molecular
orbitals, effectively infinite resolution (a limit $\gamma \rightarrow 0$) is achieved using EV-TDDFT
(Fig.~\ref{fig:W_complex_EV}).

\begin{table}
 \centering
 \caption{Metal XAS edge positions and SO splittings (in eV) for different absorption edges and compounds --
 comparison of different X2C Hamiltonian models with reference 4c data and experimental values with
 Error comparing the most accurate X2C approach with the experiment.
 Theoretical values obtained with DR-TDDFT/PBE0-$x$HF/Dyall's uDZ/uaDZ basis. The complete dataset is reported in Ref.~\citenum{Konecny2023}.
 }
 \label{tab:XAS}
 \begin{tabular}{l*{8}r}
 \hline \hline\noalign{\smallskip}
{}&                     {}&     $x$HF&    {1eX2C}          &  {(e)amfX2C}         &  {mmfX2C}  &       4C &       Exp &  Error    \\\hline
\\  
\ce{MoS4^{2-}}&    \ce{M5}&        40&   228.{\color{red}5}&                228.9 & 228.9      &    228.9 &     228.7 &   0.2     \\
              &    \ce{M4}&        40&   232.{\color{red}8}&                232.2 & 232.2      &    232.2 &     231.7 &   0.5     \\
              & $\Delta$SO&          &    {\color{red} 4.3}&                  3.3 &   3.3      &      3.3 &       3.0 &   0.3     \\[0.5cm]
\ce{WCl6}     &    \ce{L3}&        60& 10{\color{red}185.9}& 10207.3              &  10207.3   &  10207.3 &   10212.2 &  -4.9     \\
              &    \ce{L2}&        60& 115{\color{red}80.6}& 11561.{\color{red}6} &  11561.7   &  11561.7 &   11547.0 &  14.7     \\
              & $\Delta$SO&          &  13{\color{red}94.7}&  1354.{\color{red}3} &   1354.4   &   1354.4 &    1334.8 &  19.6     \\[0.5cm]
\ce{UO2(NO3)2}&    \ce{M5}&        60&  35{\color{red}39.2}&  3549.9              & 3549.9     &   3549.9 &    3549.9 &           \\
              &    \ce{M4}&        60&  37{\color{red}40.5}&  3728.0              & 3728.0     &   3728.0 &    3727.0 &   1.0     \\
              & $\Delta$SO&          &   2{\color{red}01.3}&   178.1              &  178.1     &    178.1 &     178.1 &           \\
\\
\hline \hline
\end{tabular}
\end{table}

The two linear response approaches, DR-TDDFT and EV-TDDFT, present complementary ways of
obtaining molecular properties with each method being advantageous for different applications:
DR-TDDFT calculates the spectrum directly for given frequencies while including the
contributions of all transitions, including those lying outside the frequency range, while EV-TDDFT calculates the exact values of excitation energies and response vector of the desired number of excited states. Their X2C formulation allows accurate yet efficient inclusion of relativistic effects,
both scalar and spin--orbit, with accuracy indistinguishable from 4c methods for
about a tenth of the cost.

% \subsubsection{Summary}
% Linear response TDDFT is available in \ReSpect{} with all described X2C Hamiltonians in the form
% of DR-TDDFT and EV-TDDFT and allows the calculation of the following molecular properties:
% \begin{itemize}
%     \item excitation energies, including zero-field splittings,
%     \item electronic absorption spectra,
%     \item electronic circular dichroism spectra,
%     \item frequency-dependent electric dipole polarizability,
%     \item optical rotatory dispersion,
%     \item radiative rates and radiative lifetimes using Boltzmann averaging.
% \end{itemize}
% The current list of features includes:
% \begin{itemize}
%     \item 1c(KS), 2c (amfX2C and 1eX2C), and 4c (DC) Hamiltonians,
%     \item noncollinear DFT potentials and kernels,
%     \item Tamm--Dancoff approximation for EV-TDDFT,
%     \item core--valence separation for the calculations in the X-ray region,
%     \item alternatively the possibility of applying the perturbation only to a selected MO window,
%     \item visualization of transition densities,
%     \item analysis of spectral lines in terms of MO transitions,
%     \item restart from previous calculations,
%     \item resolution-of-the-identity integral acceleration for the Coulomb term (RI-J), and
%     \item hybrid MPI/OpenMP parallelization.
% \end{itemize}

%%%%%%%%%%%%%%%%%%%%%%%%%%%%%%%%%%%%%%%%%%%%%%%%%%%%%%%%%%%%%%%%%%%%
\subsection{\label{sec:applications-QE}E: X2C Quantum-Electrodynamical DFT}

An exciting research direction that has emerged in recent years involves the control of molecular
properties through strong light--matter coupling within photonic structures. These structures,
such as optical cavities and plasmonic nanostructures, confine electromagnetic fields in a way
that leads to quantized light modes while promoting strong coupling of these photons to embedded
atoms, molecules, and materials. Under strong coupling, hybrid light–matter states called polaritons
form and alter the molecular ground and excited states. This, in turn, allows for the modulation of
energy levels, transition properties, and reaction dynamics, thereby offering a new strategy to
control chemical reaction rates, electron and energy transport, and induce new phases of
matter.~\cite{Hutchison2012, Garcia-Vidal2021}
These advances, initially driven by pioneering experiments and simplified model-based calculations,
have motivated the development of {\it ab initio} computational methods able to simultaneously capture both
the electronic structure of matter and the quantized transverse cavity photon modes.~\cite{Foley2023, Ruggenthaler2023}
One such method is quantum electrodynamical density functional theory (QEDFT) that translates
the advantages of DFT to the description of coupled light--matter systems.~\cite{Ruggenthaler2011, Ruggenthaler2014}
QEDFT has been implemented in various regimes including the description of modified
ground states in cavities~\cite{Flick2018}, and the calculation of a response to external fields
by means of the Sternheimer equation~\cite{Welakuh2022}, and real-time dynamics~\cite{Malave2022}.
However, the most common formulation is the linear response QEDFT method based on an eigenvalue
equation analogous to Eq.~\eqref{eq:EV-TDDFT}, for the calculation of polaritonically modified
spectra of molecules embedded in cavities.~\cite{Flick2019}

While most of the previous implementations of QEDFT were based on non-relativistic
description of the electronic structure, there are compelling reasons for the pursuit
of relativistic QEDFT.~\cite{Konecny2024, Thiam2024}
Multi-component relativistic methods offer a high level of precision
in describing both scalar and SO relativistic
effects. As such, they enable accurate treatment of heavy elements and processes driven by
SO coupling, whose modification has been proposed as a potential application of cavity control.
Specifically, reverse intersystem crossing proceeds via a singlet--triplet transition
that is forbidden in non-relativistic theories, and its rate depends on the relative
energies of excited singlet and triplet states that can be tuned by the strong
light--matter coupling in a cavity.~\cite{Stranius2018, Eizner2019} 
Moreover, the systems of interest for these applications contained heavy elements to
further enhance the singlet--triplet transitions by strong SO interaction.~\cite{Kena-Cohen2007, Hertzog2019}
In addition, relativistic QEDFT calculations provide a path toward answering fundamental
questions about light--matter interactions by lying between low-energy Pauli--Fierz QED
and full second-quantized relativistic QED.~\cite{Baez2014, Ruggenthaler2023}
However, 4c relativistic methods are often computationally prohibitive for many
applications of interest in QEDFT. These include calculations on large,
heavy-element-containing molecules, modeling ensembles of atoms and molecules
in cavities to capture collective effects, and performing repeated simulations
to construct two-dimensional spectra. Therefore, X2C approaches present
an attractive alternative because they have been shown to deliver near-4c accuracy
at significantly reduced computational cost in a variety of scenarios.

A coupled system of electrons and photons described within QEDFT evolves according to 
EOMs that take the form
\begin{subequations}
\begin{align}
\label{eq:el_EOM}
i\hbar \partial_t C_{\mu j}^{\text{4c}}(t)
& =
F^{\text{4c}}_{\mu\nu}\left(\mat{D}^\mathrm{4c},\vec{q}, t\right) C_{\nu j}^{\mathrm{4c}}(t)
, \\
\label{eq:ph_EOM}
\left(\frac{\partial^2}{\partial t^2} + \omega_\alpha^2 \right) q_\alpha (t) 
& = 
-\sqrt{2}\partial_t j_{\alpha}^{\mathrm{4c}}\left(\rho^{\mathrm{4c}}, q_\alpha, t \right)
.
\end{align}
\label{eq:ep_EOM}
\end{subequations}
Equation~\eqref{eq:el_EOM} describes electrons and represents a version of Eq.~\eqref{eq:x2c_eom} that
accounts for the coupling of the Fock matrix to the photon field.
Specifically, the Fock matrix $\mat{F}^{\mathrm{4c}}\left(\mat{D}^\mathrm{4c},\vec{q}, t\right)$
now contains electron--photon (ep) terms besides the familiar electronic (e) terms and the interaction with
an external probe field, taking the form
\begin{equation}
\begin{split}
\mat{F}^\mathrm{4c}(t,\vec{\mathcal{E}},\vec{q})
& =
\mat{F}^\mathrm{4c,e}\left[\mat{D}^\mathrm{4c}(t,\vec{\mathcal{E}},\vec{q})\right]
\\ & +
\mat{F}^\mathrm{4c,ep}\left[\mat{D}^\mathrm{4c}(t,\vec{\mathcal{E}},\vec{q}),\vec{q}\right]
- 
\mathcal{E}_u(t) \,\mat{P}_{\!u}^\mathrm{4c}
.
\end{split}
\end{equation}
Here, the ep term consists of the direct coupling to the cavity, electron self-energy,
and the electron--photon exchange--correlation potential
\begin{equation}
\begin{split}
F_{\mu\nu}^{\mathrm{4c},ep}\left[\mat{D}^\mathrm{4c}(t,\vec{\mathcal{E}},\vec{q}),\vec{q}\right]
=
- \sum_{\alpha=1}^{M} \omega_{\alpha} g_{\alpha} q_{\alpha}(t) \left( \vec{\epsilon}_\alpha \cdot \vec{P}^\mathrm{4c}_{\mu\nu} \right) \\
+ \sum_{\alpha=1}^{M} g_{\alpha}^2 \left(\vec{\epsilon}_\alpha \cdot \vec{P}^\mathrm{4c}_{\kappa\lambda} \right) D^{\mathrm{4c}}_{\lambda\kappa}(t) \left(\vec{\epsilon}_\alpha \cdot \vec{P}^\mathrm{4c}_{\mu\nu} \right) \\
+ \int d^{3}r\, v^{\mathrm{xc,ep}}\left[\vec{\rho}^\mathrm{4c}(\vec{r},t,\vec{\mathcal{E}},\vec{q}),\vec{q}\right] \Omega_{\mu\nu}^{\mathrm{4c}}(\vec{r})
,
\end{split}
\end{equation}
with $g_{\alpha}$ being the coupling strength, $q_{\alpha}$ the photon displacement coordinate,
$\vec{\epsilon}_\alpha$ the mode polarization, and the index $\alpha$ running over all $M$ photon modes.
The coupling strength $g_{\alpha}$ depends on the dimensions of the cavity as well as the collective
coupling effects and in QEDFT calculations is normally treated as an empirical input parameter.
Eq.~\eqref{eq:ph_EOM} accounts for the photonic degrees of freedom, defined by the displacement coordinate,
by describing their time evolution driven by the electronic and external currents.
Moreover, $\omega_\alpha$ is the frequency of mode $\alpha$, and when assuming only
electron dipole coupling the current has the form
\begin{equation}
\partial_t j_{\alpha}^{\mathrm{4c}}\left(\rho, \vec{q}, t\right)
=
\frac{\omega_\alpha g_\alpha}{\sqrt{2}} \left(\vec{\epsilon}_{\alpha} \cdot \vec{P}^\mathrm{4c}_{\mu\nu} \right) D^{\mathrm{4c}}_{\nu\mu}(t)
+ \partial_t j^{\mathrm{ext}}_{\alpha}(t)
.
\end{equation}

Equations~\eqref{eq:ep_EOM} are the starting point for the derivation of X2C-based linear response QEDFT.
They can be X2C transformed similarly as the electronic KS equation in the previous section with the
adiabatic X2C approximation analogously allowing the use of the decoupling matrix obtained from the
SCF procedure to picture change transform all relevant 4c matrices
($\mat{F}^\mathrm{4c}$, $\mat{D}^\mathrm{4c}, \mat{P}^\mathrm{4c}$).~\cite{Konecny2025}
By applying linear response theory on the transformed set of EOMs we obtain the central equation
of linear response QEDFT that is analogous to the EV-TDDFT equation while containing blocks describing
the cavity photons and the electron--photon interaction,
\begin{equation}
\label{eq:CavityCasida}
    \begin{pmatrix}
    \mat{A}^\mathrm{2c}       + \mat{\Delta}^\mathrm{2c}  &   \mat{B}^\mathrm{2c}       + \mat{\Delta}^{\mathrm{2c}'}  &   \mat{g}^\mathrm{2c}    &   \mat{g}^\mathrm{2c}    \\
    \mat{B}^{\mathrm{2c}^\ast} + \mat{\Delta}^{\mathrm{2c}\prime^\ast} &  \mat{A}^{\mathrm{2c}^\ast} + \mat{\Delta}^{\mathrm{2c}^\ast}  &   \mat{g}^{\mathrm{2c}^\ast}  &   \mat{g}^{\mathrm{2c}^\ast}  \\
    \mat{\tilde{g}}^{\mathrm{2c}^\ast}                   &  \mat{\tilde{g}}^\mathrm{2c}              & \bm{\omega}  &   \mat{0}    \\
    \mat{\tilde{g}}^{\mathrm{2c}^\ast}                   &  \mat{\tilde{g}}^\mathrm{2c}              & \mat{0}      &  \bm{\omega} \\
    \end{pmatrix}
\begin{pmatrix}
\mat{X}_n \\
\mat{Y}_n \\
\mat{M}_n \\
\mat{N}_n \\
\end{pmatrix}
=
\omega_n
\begin{pmatrix}
\mat{1} &  \mat{0} & \mat{0} &  \mat{0} \\
\mat{0} & -\mat{1} & \mat{0} &  \mat{0} \\
\mat{0} &  \mat{0} & \mat{1} &  \mat{0} \\
\mat{0} &  \mat{0} & \mat{0} & -\mat{1} \\
\end{pmatrix}
\begin{pmatrix}
\mat{X}_n \\
\mat{Y}_n \\
\mat{M}_n \\
\mat{N}_n \\
\end{pmatrix}
.
\end{equation}
The new terms in Eq.~\eqref{eq:CavityCasida} compared to Eq.~\eqref{eq:EV-TDDFT} are the electron--photon
and photon--electron coupling blocks
\begin{equation}
    g_{ai,\alpha} = \tilde{g}_{\alpha,ai} =
    \sqrt{\frac{\omega_\alpha}{2}}\, g_\alpha \, \vec{\epsilon}_\alpha \cdot \vec{\tilde{P}}^\mathrm{2c}_{ai}
    ,
\end{equation}
and self-energy terms
\begin{subequations}
\label{eq:SelfEnergy}
\begin{align}
    \Delta_{ai,bj}^\mathrm{2c}  & = \frac{g_\alpha^2}{2} \left(\vec{\tilde{P}}^\mathrm{2c}_{ai} \cdot \vec{\epsilon}_\alpha \right) \left(\vec{\tilde{P}}^\mathrm{2c}_{jb} \cdot \boldsymbol{\epsilon}_\alpha \right) , \\
    \Delta^{\mathrm{2c}'}_{ai,bj} & = \frac{g_\alpha^2}{2} \left( \vec{\tilde{P}}^\mathrm{2c}_{ai} \cdot \vec{\epsilon}_\alpha \right) \left( \vec{\tilde{P}}^\mathrm{2c}_{bj} \cdot \boldsymbol{\epsilon}_\alpha \right) ,
\end{align}
\end{subequations}
in the coupling matrix, and the photon creation and annihilation amplitudes
($\mat{M}_n$, $\mat{N}_n$ respectively) in the response vector.
In addition, the equation can be approximated in several ways in either the electronic or photonic
subsystem such as by introducing TDA as in EV-TDDFT or by similarly neglecting the contribution of
photon deexcitation (annihaltion) terms, i.e. the rotating wave approximation (RWA), neglecting
the self-energy terms, i.e. the Rabi approximation, and performing both, i.e. the Jaynes--Cummings
(JC) approximation.~\cite{Yang2021} These approximate equations are available in \ReSpect{}
with all X2C Hamiltonians.

The solutions of Eq.~\eqref{eq:CavityCasida} are the eigenvalues $\omega_n$ corresponding
to excitation energies of the coupled light-matter system
and the eigenvectors that contain the electronic excitation and deexcitation amplitudes
$\mathbf{X}_n$ and $\mathbf{Y}_n$ and the photonic creation and annihilation amplitudes
$\mathbf{M}_n$ and $\mathbf{N}_n$.
The linear response equation~\eqref{eq:CavityCasida} is formulated in the basis of the ground state
molecular orbitals for the electron variables and photon modes for light.
In applications involving optical cavities---particularly ideal cavities---the number of photon
modes considered is typically very low, often limited to a single mode. As a result, the equation is dominated by
the electronic subsystem, which allows the use of the iterative subspace solver developed for EV-TDDFT,
with an appropriate extension to handle the photon blocks.

Equation~\eqref{eq:CavityCasida} is solved for the excitation energies $\omega_n$ of the coupled light--matter
system and their corresponding transition vectors. These can be used to calculate an electronic absorption
spectrum of a molecule embedded in an optical cavity using Eqs.~\eqref{eq:rspFunction}, \eqref{eq:Polarizability-EV},
and \eqref{eq:transitionDipoleEV} where the photonic parts $\mathbf{M}_n$ and $\mathbf{N}_n$ of the electron--photon
transition vector do not enter the formulas because the electric dipole moment operator does not act in the
photon space. Such a spectrum provides information about the changes in the excited state manifold
resulting from the strong coupling and serves as the starting point for assessing the modification
of chemical and physical properties of the molecule.

\begin{figure}[h]
\centering
\caption{Spectra of a mercury porphyrin complex in an optical cavity.
}
  \begin{subfigure}{0.49\textwidth}
    \includegraphics[width=\textwidth]{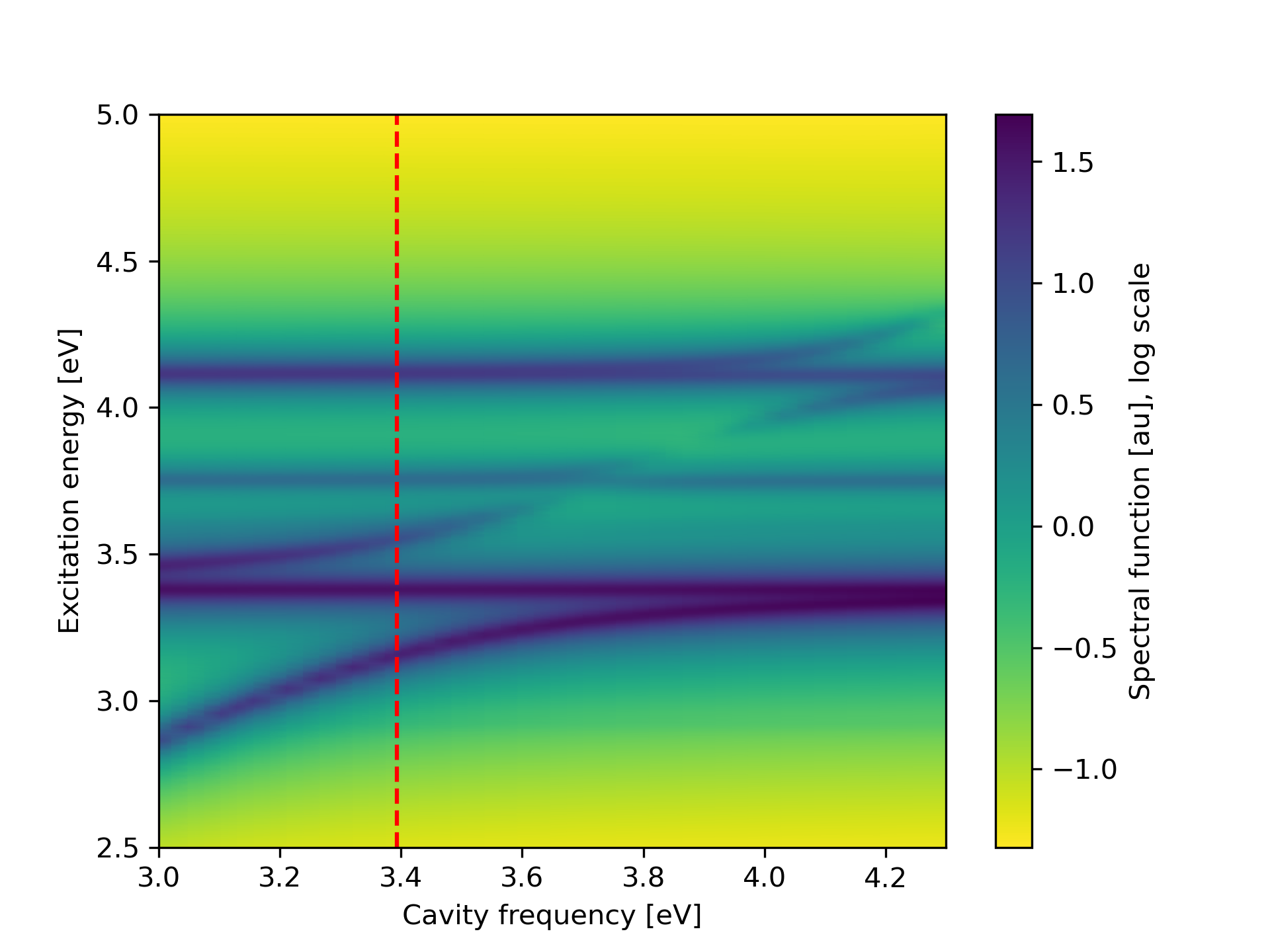}
    \caption{2D spectrum}
    \label{fig:HgP_2D}
  \end{subfigure}
  \begin{subfigure}{0.49\textwidth}
    \includegraphics[width=\textwidth]{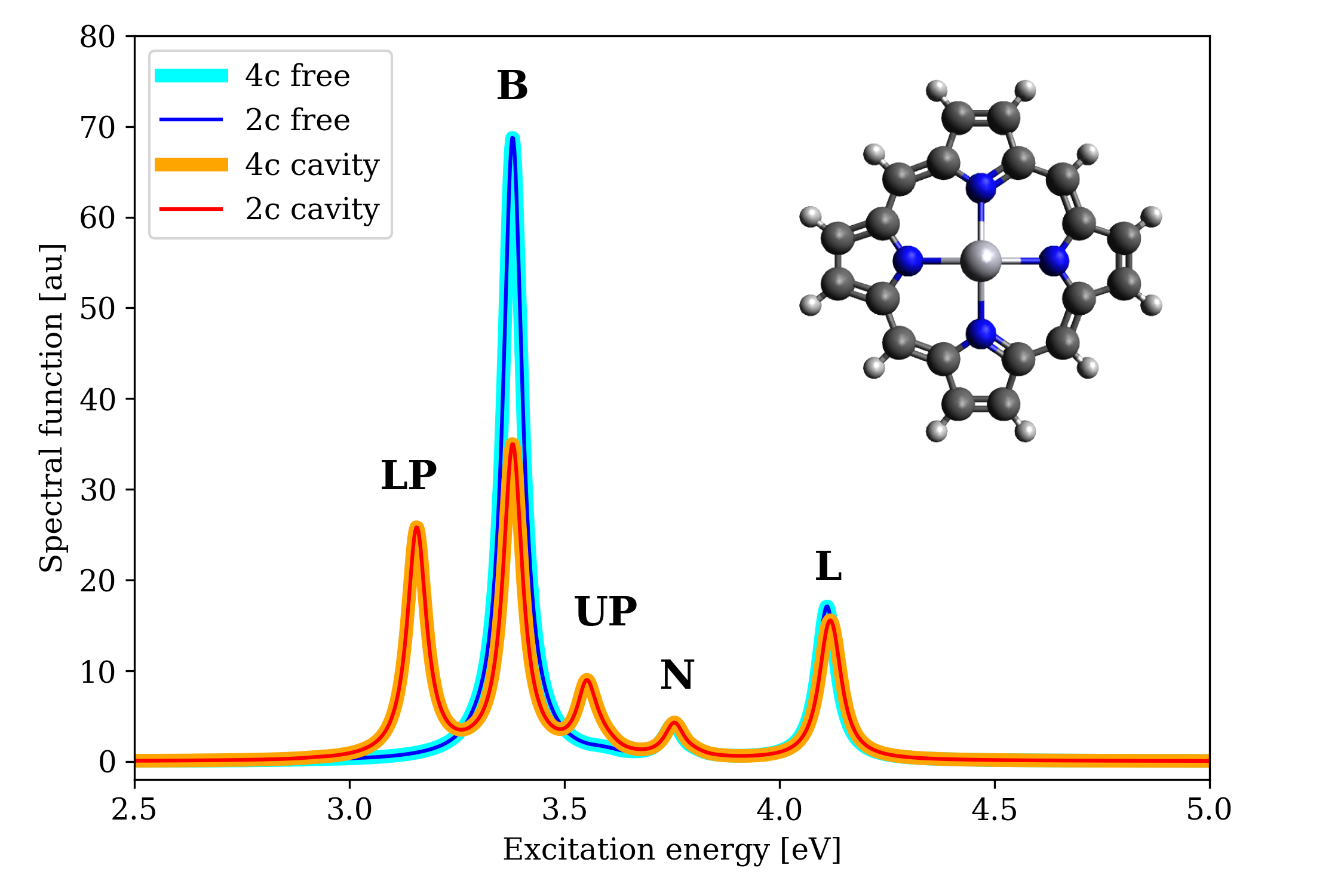}
    \caption{Spectra of free molecule and in cavity in resonance with the B line
    [red dashed line in panel (a)].}
    \label{fig:HgP_res}
  \end{subfigure}
  \label{fig:HgP_in_cavity}
\end{figure}

As an illustration, let us investigate a mercury porphyrin embedded in and ideal Fabry--P\'{e}rot cavity
consisting of two parallel mirrors and possessing only a single photon mode. The frequency of the mode
can be tuned by varying the distance between the mirrors. The molecule lies in the $yz$ plane while
the cavity mode is polarized along the $z$ axis and we chose the coupling strength $g_\alpha = 0.01$.
The spectrum of the uncoupled system (shown light and dark blue lines for 4c and amfX2C level of theory
in Fig.~\ref{fig:HgP_res}) is dominated by three lines conventionally called B, N, and L, each consisting
of two degenerate states with perpendicular transition moments. We used the broadening parameter
$\gamma = \unit[0.027]{eV}$ to obtain the band spectra.  By varying the cavity frequency and recording
the corresponding absorption spectra we can construct a 2D spectrum such as depicted in Fig.~\ref{fig:HgP_2D}
(using the same broadening). As the frequency of the cavity mode increases, it hybridizes with different
excited states to form polaritonic states. Fig.~\ref{fig:HgP_res} depicts one such absorption
spectrum where the cavity frequency was set to resonance with the B line and corresponds to the
red dashed line in Fig.~\ref{fig:HgP_2D}. The spectrum in cavity contains two noticeable peaks
corresponding to the lower polariton (LP) and the upper polariton (UP). These result from the
mixing of the cavity excitation with one component of the B line whose transition moment is parallel
with the cavity mode polarization. The second degenerate state from the B line with a perpendicular
transition moment remains uncoupled resulting in a persisting signal at the original B line energy
albeit with half the intensity.

In Fig.~\ref{fig:HgP_res} we also see that the amfX2C-based QEDFT calculation exactly reproduces
the reference 4c spectrum. However, since the amfX2C calculation is more than 5-times faster,
amfX2C QEDFT enables the computation of 2D spectra that would be overly expensive in the 4c regime
for such a system.
The spectrum shown in Fig.~\ref{fig:HgP_2D} consists of 31 individual QEDFT absorption spectra.
Owing to the restart scheme implemented in \ReSpect{}, the QEDFT calculations were performed starting
from the initial guess provided by EV-TDDFT results, thus requiring only a few iterations and each
taking about 1/10th of the time required for the reference EV-TDDFT calculation.
Let us also note that while the spectrum is dominated by three main lines (more precisely, the number
of excited states with oscillator strength higher than $10^{-5}$, $10^{-3}$, and $10^{-2}$ is 39, 13,
and 6, respectively), the use of a few-level model description, e.g. a four-level model consisting of
the ground state and three excited states, would not be more advantageous compared to QEDFT simulations.
To parametrize such model one would need to run a reference EV-TDDFT calculation for excitation
energies and transition moments required by typical models. However, such a calculation already presents
a significant computational investment, as it would still need to be run for many excited states since
the excited states are typically doubly degenerate, and the L line corresponds to far lying excited
states no.\ 81 and 82 due to the high number of dark states. It would then have to be followed
by the model calculations that in the standard parametrization neglect the dipole self-energy and the coupling
between the excited states and may need refining, unlike the first principles QEDFT.
Therefore, linear response QEDFT based on X2C Hamiltonians presents an efficient first-principles
approach to calculating properties of matter in optical cavities applicable even to large molecules
with dense spectra.

% \subsubsection{Summary}
% Linear response QEDFT is available in \ReSpect{} with all described X2C Hamiltonians in the form
% of coupled electron--photon eigenvalue equation and allows the calculation of the following properties
% of molecules embedded in optical cavities:
% \begin{itemize}
%     \item excitation energies,
%     \item electronic absorption spectra
% \end{itemize}
% The current list of features includes:
% \begin{itemize}
%     \item 1c(KS), 2c (amfX2C and 1eX2C), and 4c (DC) Hamiltonians,
%     \item noncollinear DFT potentials and kernels,
%     \item Tamm--Dancoff approximation for the electronic subsystem,
%     \item approximations for the photon subsystem including the Jaynes--Cummings, Rabi,
%           and rotating wave approximation,
%     \item visualization of transition densities,
%     \item restart from previous calculations,
%     \item resolution-of-the-identity integral acceleration for the Coulomb term (RI-J), and
%     \item hybrid MPI/OpenMP parallelization.
% \end{itemize}

%%%%%%%%%%%%%%%%%%%%%%%%%%%%%%%%%%%%%%%%%%%%%%%%%%%%%%%%%%%%%%%%%%%%%%%%%%%%%%%%%%%%%%%%%%%%%%%%%%%%%%%%%
\section{Summary and Outlook}

Since the initial publication detailing the theoretical foundations and computational implementation of our relativistic density functional theory program \ReSpect{}, the code has continued to grow in support of its twin goals: providing accessible tools for the simulation of spectroscopic processes, and enabling exploration of emerging research areas--while treating relativistic effects, particularly spin–orbit interactions, in a fully variational manner.

The primary focus of development in recent years has been on exact two-component (X2C) Hamiltonian models that extend beyond the conventional one-electron X2C approach by incorporating two-electron picture-change corrections in a simple, computationally efficient, and numerically accurate manner.
This manuscript summarizes the essential theoretical foundations of two distinct atomic mean-field X2C models--developed within \ReSpect{}--namely, amfX2C and its extended variant, eamfX2C. These models offer accuracy comparable to fully relativistic four-component calculations, but at a significantly reduced computational cost. Moreover, the implementation of (e)amfX2C models has enabled the simulation of more complex phenomena, such as time-resolved pump–probe spectroscopies and cavity-induced modifications of molecular properties, which would otherwise be computationally prohibitive using full four-component methods. Nevertheless, fully relativistic four-component approaches remain available in \ReSpect{} and continue to be developed to provide reliable reference data.

\ReSpect{} is a continuously evolving code, with new techniques and improvements developed and published annually to expand the range of relativistic calculations available to interested users. In addition to providing state-of-the-art quantum chemical methods, we also offer tools for post-processing and visualization, along with a comprehensive manual and worked examples for all available molecular properties. These resources are accessible on our website at www.respectprogram.org. The website also provides links to download the compiled program free of charge for researchers in chemistry, physics, materials science, and other disciplines who wish to explore relativistic effects.

%%%%%%%%%%%%%%%%%%%%%%%%%%%%%%%%%%%%%%%%%%%%%%%%%%%%%%%%%%%%%%%%%%%%%
%% The "Acknowledgement" section can be given in all manuscript
%% classes.  This should be given within the "acknowledgement"
%% environment, which will make the correct section or running title.
%%%%%%%%%%%%%%%%%%%%%%%%%%%%%%%%%%%%%%%%%%%%%%%%%%%%%%%%%%%%%%%%%%%%%
\begin{acknowledgement}

MR gratefully acknowledges Stefan Knecht, Hans Jørgen Aagaard Jensen, and Trond Saue for the fruitful collaboration on the development of the (e)amfX2C models.
We acknowledge the support received from the Research Council of Norway through
a Centre of Excellence Grant (no. 262695), % Hylleraas 
research grant (no. 315822), %KR and MR
mobility grants (no. 301864 and no. 314814). %MK, LK. 
T.M. acknowledges the support from the Marie
Sklodowska-Curie individual postdoctoral fellowship (grant no. 101152113). In
addition, M.R. acknowledges funding from the European Union’s Horizon 2020
research and innovation program under the Marie Sklodowska-Curie Grant
Agreement no. 945478 (SASPRO2), the Slovak Research and Development Agency
(grant no. APVV-22-0488), VEGA no. 1/0670/24 and 2/0118/25, and the EU NextGenerationEU through the Recovery and Resilience Plan for Slovakia under the project No. 09I05-03-V02-00034.
We thank Sigma2 - The National Infrastructure for High Performance Computing and Data Storage in Norway, grant
no. NN14654K, and EuroHPC regular access grant no. EU-25-8 for the computational resources. 

\end{acknowledgement}

%%%%%%%%%%%%%%%%%%%%%%%%%%%%%%%%%%%%%%%%%%%%%%%%%%%%%%%%%%%%%%%%%%%%%
%% The same is true for Supporting Information, which should use the
%% suppinfo environment.
%%%%%%%%%%%%%%%%%%%%%%%%%%%%%%%%%%%%%%%%%%%%%%%%%%%%%%%%%%%%%%%%%%%%%
%\begin{suppinfo}
%
%This will usually read something like: ``Experimental procedures and
%characterization data for all new compounds. The class will
%automatically add a sentence pointing to the information on-line:
%

%\end{suppinfo}

%%%%%%%%%%%%%%%%%%%%%%%%%%%%%%%%%%%%%%%%%%%%%%%%%%%%%%%%%%%%%%%%%%%%%
%% The appropriate \bibliography command should be placed here.
%% Notice that the class file automatically sets \bibliographystyle
%% and also names the section correctly.
%%%%%%%%%%%%%%%%%%%%%%%%%%%%%%%%%%%%%%%%%%%%%%%%%%%%%%%%%%%%%%%%%%%%%
\bibliography{article}

\end{document}